\documentclass[namedreferences]{solarphysics}
\usepackage[hyperref,optionalrh,solaromanenum]{spr-sola-addons} 
\pdfoutput=1 
\usepackage{cleveref}
\usepackage{epsfig}     
\usepackage{graphicx}    
\usepackage{color}       
\usepackage{color, colortbl}
\usepackage{xcolor}
  
\usepackage{paralist}
\usepackage[shortlabels]{enumitem}
\usepackage{float}
\usepackage{verbatim}
\chardef\us=`\_

  \newcommand{\kms}{km~$\rm s^{-1}$}
  
  \newcommand{\rsun}{$R_\odot$}

  \newcommand{\sws}{$\rm{V}_{\rm{sw}}$}

\usepackage[printwatermark]{xwatermark}
\usepackage{eso-pic}
\usepackage{tikz}
\newsavebox\mybox
\savebox\mybox{\tikz[color=gray,opacity=0.3]\node{Preprint: Solar Physics};}
\newwatermark*[pages=2-\lastdocpage,
  angle=60,
  scale=6,
  xpos=-5,
  ypos=15
]{\usebox\mybox}

\begin{document}
  \begin{article}
    \begin{opening}
      \title{Uncertainty Estimates of Solar Wind Prediction 
             using HMI Photospheric Vector and Spatial Standard 
             Deviation Synoptic Maps}
%
%
%
\author[addressref={aff1},corref,email={bala.poduval@unh.edu}]{\inits{B.P}\fnm{B.}\lnm{Poduval}}
\author[addressref={aff2},corref,email={gpetrie@nso.edu}]{\inits{G.P}\fnm{G.}\lnm{Petrie}}
\author[addressref={aff2},corref,email={lbertello@nso.edu}]{\inits{LB}\fnm{L.}\lnm{Bertello}}
%
%
      \runningauthor{Poduval et al.}
      \runningtitle{Uncertainty Estimates of Solar Wind Prediction}
%
%
%
      \address[id={aff1}]{University of New Hampshire, Durham, NH 03824}
      \address[id={aff2}]{National Solar Observatory, Boulder, CO 80303}
%
%

\begin{abstract}
The solar wind prediction is based on the Wang \& Sheeley empirical 
relationship between the solar wind speed observed at 1~AU and the 
rate of magnetic flux tube expansion (FTE) between the photosphere 
and the inner corona, where FTE is computed by coronal models 
(e.g., Current Sheet Source Surface (CSSS) and Potential Field Source 
Surface: (PFSS)). These models take the photospheric flux density 
synoptic maps as their inner boundary conditions to extrapolate the 
photospheric magnetic fields to deduce the coronal and the heliospheric 
magnetic field (HMF) configuration. These synoptic maps are among the 
most widely-used of all solar magnetic data products and therefore, 
the uncertainties in the model predictions that are caused by the 
uncertainties in the synoptic maps are worthy of study. However, an 
estimate of the uncertainties in the construction of these synoptic 
maps were not available until recently when Bertello et~al. 
(\textit{Solar Physics}, {\bf 289}, 2014) obtained the spatial 
standard deviation synoptic maps. For each photospheric synoptic maps, 
they obtained 98~Monte-Carlo realizations of the spatial standard 
deviation maps. \\[5pt]
In this paper, we present an estimate of uncertainties in the solar 
wind speed predicted at 1~AU by the CSSS model due to the uncertainties 
in the photospheric flux density synoptic maps. We also present a 
comparison of the coronal hole (CH) locations predicted by the models 
with the \textit{STEREO}/SECCHI EUV synoptic maps. For the present 
study, we used the HMI vector/longitudinal photospheric synoptic maps
and the corresponding spatial standard deviation maps. In order to quantify 
the extent of the uncertainties involved, we compared the predicted 
speeds with the OMNI solar wind data during the same period (taking the 
solar wind transit time into account) and obtained the root mean square 
error (RMSE) between them. To illustrate the significance of the 
uncertainty estimate in the solar wind prediction, we carried out the 
analysis for three Carrington rotations, CR~2102 (3~-- 30~October, 2010), 
CR~2137 (14~May --~11~June, 2013) and CR~2160 (1~-- 28~February, 2015), 
which fall within the extended minimum, the late-ascending and the 
early-descending phases, respectively, of the solar cycle~24.\\[5pt]
The uncertainty estimate is critical information necessary for the 
current and future efforts of improving the solar wind prediction 
accuracies. In this paper, our aim is not to produce improved 
solar-wind predictions, nor to redesign the synoptic maps or models. 
Instead, we take the existing, widely-used synoptic maps and models, 
and quantify the propagation of uncertainties from the synoptic maps 
(derived straight from the statistics feeding each synoptic map pixel) 
to the models. 
\end{abstract}

%
%
\keywords{Solar wind, Space weather, Uncertainty estimate, 
  Magnetic fields, Synoptic map, Corona}
\end{opening}
%
%
\section{Background}
         \label{sec:intro} 

\begin{figure}  
  \vspace{-1.5in}
  \centerline{\includegraphics[width=1.0\textwidth,clip=]{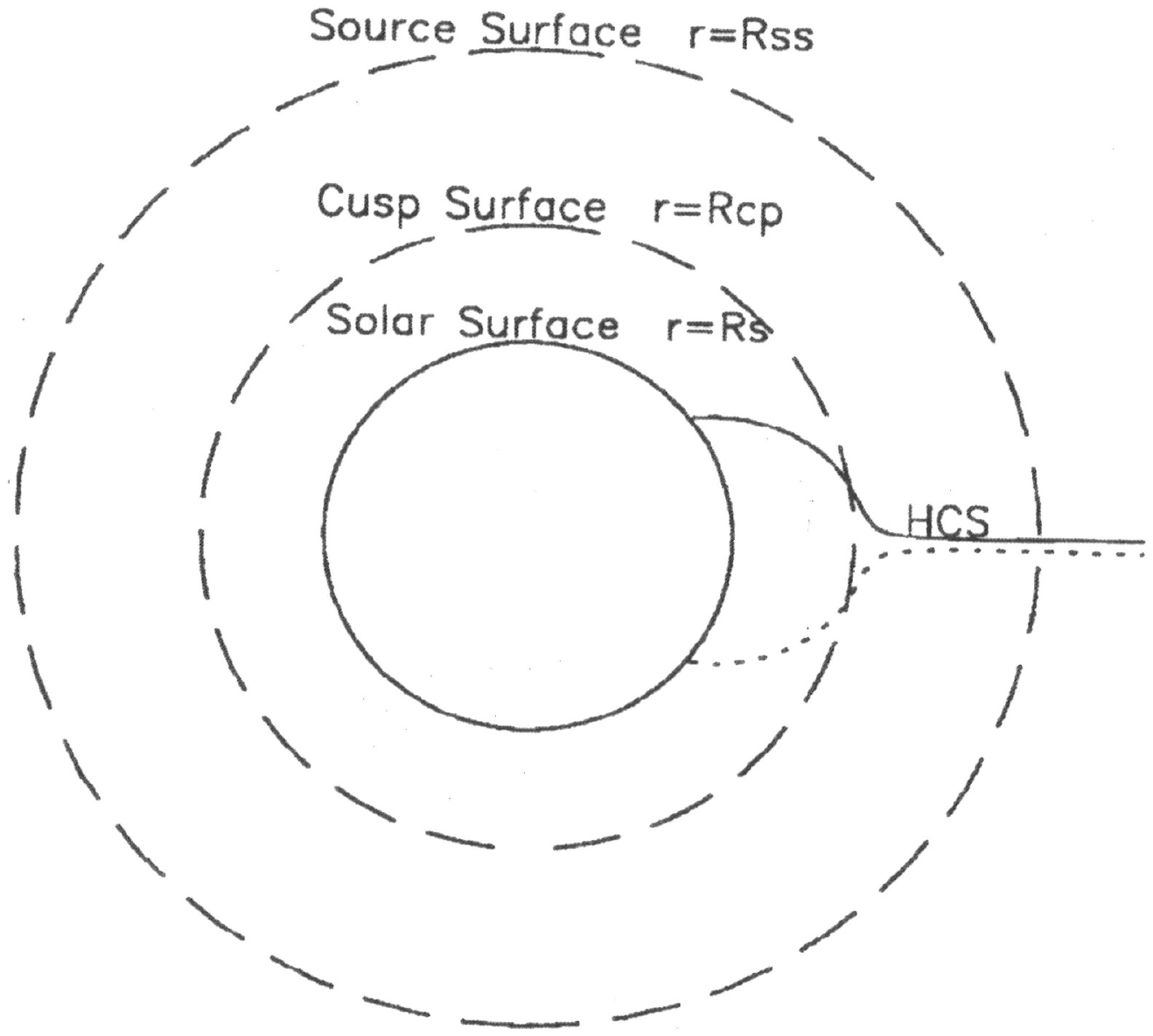}
             }
  \vspace{-1.5in}                 
  \caption{Geometry of the CSSS Model \citep{zha95a}. The 
           locations of the cusp surface, corresponding to the 
           locus of cusp points of helmet streamers, ($R_{cp}$) 
           and the source surface ($R_{ss}$) are free parameters 
           of the model; $R_s$ is the radius of the photosphere.
          }
          \label{fig:geometry}
\end{figure}

The solar magnetic field plays a significant role in controlling the 
solar wind outflow, and it is well--known that the solar wind behavior 
is greatly influenced by the shape of individual bundles of magnetic 
field lines 
\citep{zir77a,zir77b,lev77,wan90a,fis99a,fis99b,koj07a,cra13,kri73,ril15}. 
However, the mechanisms that give rise to the observed slow and fast 
solar wind streams -- the two distinct components with differing physical 
properties -- are still not understood satisfactorily. This is mainly due 
to lack of direct measurements of near--Sun solar wind properties, except 
for the Parker Solar Probe measurements released in November 2019, and 
limited observation of coronal magnetic field 
\cite[e.~g. ][]{bak13,rac13,dov11}. Much of our current understanding is 
based on, in addition to the photospheric observations of the Sun and 
near--Earth solar wind measurements, the magnetohydrodynamic (MHD) 
models and the magnetostatic models of the corona such as the Potential 
Field Source Surface \citep[PFSS:][]{alt69,sch69} and the Current Sheet 
Source Surface \citep[CSSS: Figure~\ref{fig:geometry}; ][]{zha95a} models. 

The current solar wind prediction scheme is built on the inverse 
correlation (the well-known Wang \& Sheeley empirical relation) between 
the rate of expansion of the magnetic flux tubes (FTEs: described in 
detail below) in the inner corona (below 2.5~\rsun), computed using 
coronal extrapolation models, such as CSSS and PFSS models, and the 
observed solar wind at 1~AU. These models extrapolate the observed 
photospheric magnetic fields to deduce the coronal and the heliospheric 
magnetic field (HMF) configuration, making use of the photospheric flux 
density synoptic maps (top panel in Figure~\ref{fig:hmi_maps})
constructed from the magnetograms measured by 
ground-based and spacecraft observatories as the lower boundary conditions. 
These synoptic maps are among the most widely-used of all solar magnetic 
data products. Moreover, as well-known, the quality and reliability of 
these synoptic maps are critical to the accuracy of solar wind predictions 
(speed \& magnetic field), determination of locations of coronal holes 
(CH) and the magnetic neutral line (NL), and the global structure and 
properties of many other solar and heliospheric phenomena 
\citep[e.~g.][]{ril14,arg00}. Therefore, the uncertainties in the model 
predictions that are caused by the uncertainties in the synoptic maps 
are worthy of study. An estimate of the uncertainties in the construction 
of these synoptic maps were not available until recently when \citet{ber14} 
obtained the spatial standard deviation synoptic maps (bottom panel in 
Figure~\ref{fig:hmi_maps}). For each photospheric synoptic maps, there 
are 98~Monte-Carlo realizations of the spatial standard deviation maps. 
Instrumental noise in the magnetic field measurements, the spatial 
variance arising from the magnetic flux distribution and the temporal 
evolution of magnetic field contribute to these uncertainties. 
\citet{ber14} have shown that these uncertainties led to significant 
differences in the CH and NL locations obtained using the PFSS model. 

In the theory of solar wind by \citet{par58}, there exists a reference 
height above the photosphere beyond which the magnetic field is 
dominated by thermal pressure and inertial force of the expanding solar 
wind. To mimic the effects of the solar wind expansion on the field, an 
equipotential upper boundary above the photosphere (a source surface 
with radius $R_{ss}$) has been introduced in the PFSS model, forcing 
the field to be open and radial on this surface \citep{alt69,sch69,sch71}. 
Following \citet{hoe84}, it is customary to place the source surface 
$R_{ss}$ at 2.5~\rsun$\;$although other choices have led to more 
successful reconstructions of coronal structures during different phases 
of the solar cycle \citep{lee11,ard14}. The lower boundary has been 
identified with the photosphere. Assuming the region between these two 
boundaries to be current--free, the PFSS model solution is uniquely 
determined in the domain \rsun$\; \le r \le R_{ss}$. The PFSS model 
is the simplest and the most widely--used global coronal model. The 
open--field footpoints obtained by PFSS model can be compared with 
observed coronal holes while the neutral lines at the outer boundary 
can be compared with observed streamer locations. It serves as a useful 
reference for more sophisticated models. 
     
Though still debated, the reference height where the plasma takes over the 
magnetic force, is considered to be $\sim 10-20$~\rsun$\;$\citep{sch71,zha10}. 
Moreover, though magnetic field lines are open above a height around 
2.5~\rsun, the coronal field is not radial, as evident from many 
observations, until farther out in the corona. In the CSSS model, these 
two heights are represented by a \textit {cusp surface} and a \textit 
{source surface} as shown in Figure~\ref{fig:geometry}, typically placed 
at 2.5 and 15~\rsun, respectively. Moreover, the field lines are allowed to 
be non--radial between the cusp surface and the source surface though they 
are all open at the cusp surface \citep{zha95a}. This is consistent with 
observations of the Large Angle and Spectrometric Coronagraph (LASCO) 
instrument on board the Solar and Heliospheric Observatory (\textit{SOHO}), 
which revealed that field lines, except near the magnetic neutral line, are 
non--radial until further out in the corona \citep[e. g.,][]{zha02,wan96b}. 
Further, the corona is not strictly current--free as evidenced by the 
numerous structures and features of the corona \citep{zha95a,hun72,pne71}. 
The volume and sheet currents in the lower corona couple with the 
magnetohydrodynamic forces to produce the distension of coronal magnetic 
field lines into an open configuration -- the solar wind, the helmet 
streamers and the coronal holes are all evidence of the existence of 
these currents in the corona. Coronal currents, being complex and widely 
distributed, are modeled as volume currents while the heliospheric current 
sheet is modeled as a sheet current \citep{zha95a,hun72,pne71}, maintaining 
the total pressure balance between regions of high and low plasma density. 
The CSSS model incorporates volume and sheet currents, based on the 
analytical solutions obtained by \citet{bog86} for a corona in static 
equilibrium, assuming that the electric currents are flowing horizontally 
everywhere. The lower boundary is taken as the photosphere as in the 
PFSS model. 

%
%
%
%

Owing to the treatment of electric currents in the model and the particular 
geometry, the CSSS model is considered to depict a more realistic scenario 
of the solar corona than the PFSS model. Particularly, based on the fact 
that the magnetic field lines are allowed to be non--radial between the cusp 
surface and the source surface, the CSSS model is expected to map coronal 
features or the coronal sources of fast/slow wind back to the photosphere 
with greater accuracy than the PFSS model. The CSSS model has been shown 
to predict \sws$\;$and HMF with greater accuracy than the PFSS model 
\citep{zha95a,pod14,pod16}. Also, \citet{zha02} used it to map the radial 
evolution of helmet streamers and found that their results were in general 
agreement with observations. Moreover, \citet{sch06} have shown that the 
Sun's open magnetic flux computed using CSSS model is more accurate than 
that computed using PFSS model. Further, \citet{dun05} and \citet{jac15} 
found that the coronal field and the north--south components of HMF computed 
using the CSSS model matched reasonably well with spacecraft observations. 

\citet{lev77} demonstrated that the observed solar wind speed (\sws) near 
the Earth's orbit inversely correlates with the rate of expansion of 
magnetic flux tubes (FTE) between the photosphere and the source surface 
(in the PFSS model). Mathematically,
\begin{equation}                     \label{eq:fte}
      FTE = \left (\frac {R_\odot}{R_{ss}}\right )^2
                          \frac {B_r(phot)}{B_r(ss)}
\end{equation}
\noindent where, $B_r(phot)$ and $B_r(ss)$ are the radial component of 
magnetic fields at the photosphere and the source surface, and 
\rsun$\;$and $R_{ss}$ are the respective radii. Extending this idea, Wang 
and Sheeley (WS) established an empirical relationship between FTE and 
\sws$\;$as shown in Table~\ref{tab:speedfte} 
\citep{wan90a,wan93,wan95,wan96a,wan97}. This empirical relationship forms 
the basis of the current solar wind prediction technique 
\citep[WSA: Wang--Sheeley--Arge model][]{arg00}) and is the most widely 
used coronal property to compute 
\sws$\;$\citep{arg98,owe05,mcg08,piz11,jia11,jan14,pod14,pod16}. 

%
%
\begin{table}
  \begin{tabular}{|c|c|}    
  \hline   
    Speed ($\mbox{km s}^1$) &  FTE  \\
  \hline
    $ > 750 $               &   $< 4.5 $   \\
    $ 650 - 750 $           &   $4.5 - 8$  \\
    $ 550 - 650 $           &   $8 - 10$   \\
    $ 450 - 550 $           &   $10 - 20$  \\
    $  < 450 $              &   $ > 20$     \\  [2pt]
  \hline
  \end{tabular}
  \caption{The empirical relationship between 
           SWS and FTE established by Wang 
           and Sheeley \citep{wan95,wan97}. 
          }
          \label{tab:speedfte}
  \vspace{-26pt}
\end{table}
%

The causes of the discrepancies between the current predictions of space 
weather events such as the arrival times of CMEs and high speed streams 
(HSS) at Earth, and the actual observed times \citep[e.g.][]{piz11} remain 
to be investigated in detail using novel approaches for better accuracy 
of these predictions. One critical but unavailable information in these 
predictions is an estimate of uncertainties. In this paper, we present 
the results of an investigation of how the uncertainties in the construction 
of photospheric synoptic maps affect solar wind prediction. For this, we 
obtained an estimate of the uncertainties in the prediction of \sws$\;$as 
the root mean square error (RMSE) between the predicted and the in situ 
observations (from the OMNI database) of solar wind speed. Further, we 
compared the predicted CH and NL locations with those derived from the EUV 
data the Sun Earth Connection Coronal and Heliospheric Investigation 
(SECCHI) telescopes on board the Solar TErrestrial RElations Observatory 
\textit{(STEREO)} to present the extent of variations in the computed CH 
and NL locations in relation to the observations. As a first step, we 
selected three Carrington rotations representing the minimum 
(2010, 3--30 October: CR2102), the late-ascending (2013, 14~May -- 11~June: 
CR~2137) and the early-descending (2015, 1--28 February: CR~2160) phases 
of the solar cycle, in an attempt to explore and present the significance 
of uncertainty estimate in the solar wind prediction. We expect the 
present exploratory work will lay the foundation for future comprehensive 
analyses. 

In \S~\ref{sec:data}, we present a description of the HMI vector synoptic 
maps and the derived standard deviation maps, and all other data used 
for the analysis in this paper. The method we adopted and the results are 
described in \S~\ref{sec:method} and a discussion of the results is presented 
in \S~\ref{sec:results}, followed by our concluding remarks in 
\S~\ref{sec:discussion}. 

%
%
\section{PERIOD OF STUDY, DATA \& METRICS OF ACCURACY}
              \label{sec:data}

\begin{figure}   
  \centerline{\includegraphics[width=1.0\textwidth,clip=]{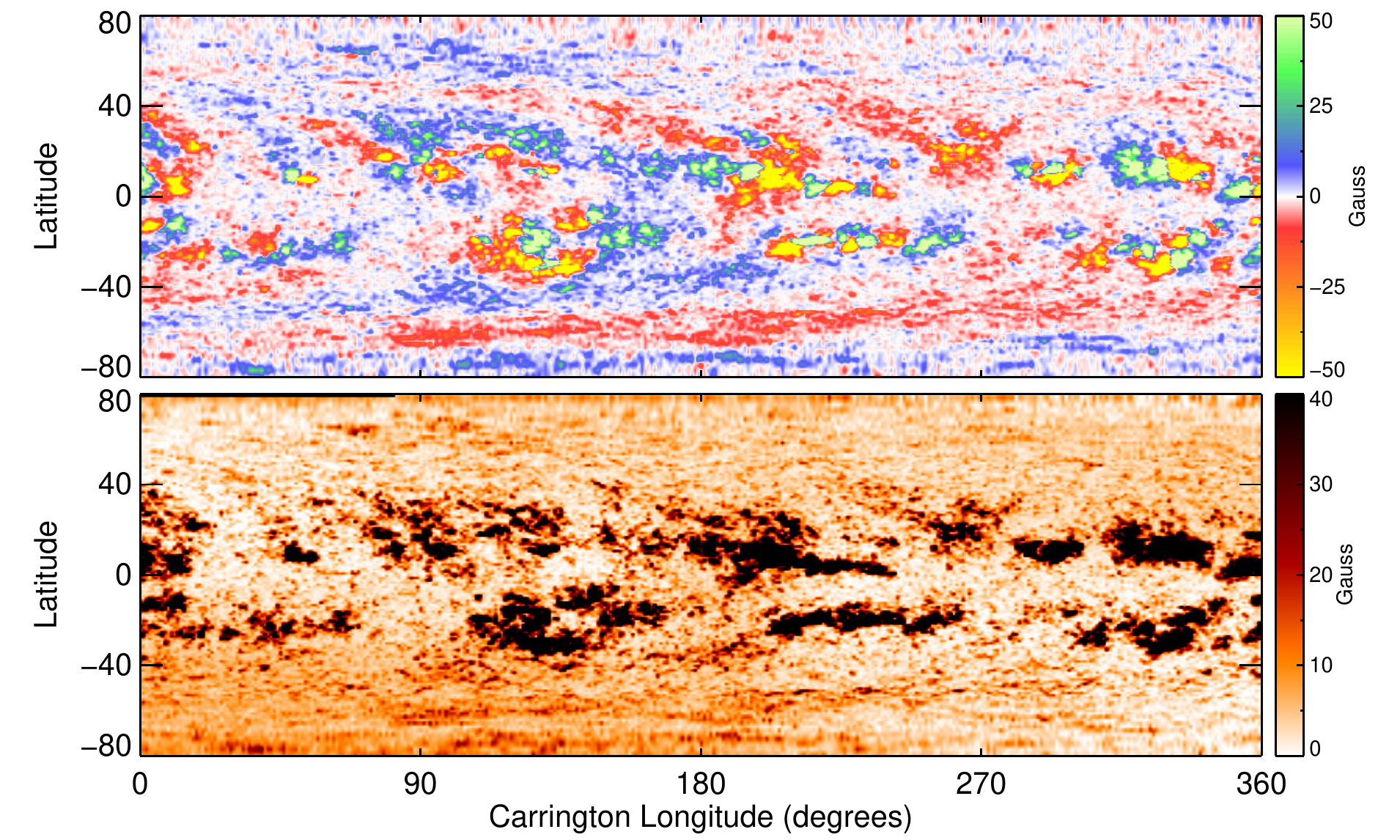}
             }
  \caption{Photospheric magnetic flux density distribution (top) 
           and corresponding standard deviation map (bottom) for 
           CR~2137. Both charts were computed using the radial 
           component of the available inverted HMI full--disk 
           vector magnetograms, following the procedure described 
           in \citep{ber14}. The top image has been scaled between 
           $\pm$50~gauss to better show the distribution of the 
           weak magnetic flux density field across the map. The 
           actual flux density distribution, however, covers a much 
           larger range of field values -- up to several hundred 
           gauss.
          }
          \label{fig:hmi_maps}
\end{figure}

The HMI synoptic maps are available from CR~2096 (since 6~May, 2010) 
and the spatial standard deviation synoptic maps from CR~2006 
(29~January -- 24~February, 2005) till present. We used the regular 
vector/longitudinal synoptic maps (processed by the NSO pipeline: Synoptic 
Maps based on \textit{SDO}/HMI observations, 
\url{https://solis.nso.edu/0/vsm/vsm\_maps.php}) and the corresponding 
ensemble of flux density synoptic maps consisting of~98 Monte-Carlo 
realizations of the spatial standard deviation maps (described in 
\S~\ref{sub:variance} and depicted in Figure~\ref{fig:hmi_maps}) for 
three Carrington rotations. For simplicity, we refer to these 98 synoptic 
maps as ``MCRs" and the regular magnetic flux density synoptic maps 
as ``HMI synoptic maps" for the remainder of this paper.

We selected CR~2102 (3~--~30~October 2010) which falls within the 
extended minimum phase, CR~2137 (14~May -- 11~June, 2013), during
the late-ascending phase and CR~2160 (1~--~28 February, 2015)
during early-descending phase of the solar cycle~24. During 
rotations CR~2102 and 2160, there were no Interplanetary Coronal 
Mass Ejections \citep[ICMEs:][]{can03,ric10} reported 
\url{http://www.srl.caltech.edu/ACE/ASC/DATA/level3/icmetable2.htm}; 
Courtesy Dr.~Ian Richardson) while in CR~2137, there were a few
solar flares and CMEs observed. The PFSS and CSSS models are 
static models and therefore, do not incorporate the effects of
transients such as solar flares and CMEs.    

The CSSS computation for a single Carrington rotation takes
about 8~hours on a laptop/desktop. For each Carrington rotation
selected for the present study, there are 98~MCRs (equivalent to 
98 Carrington rotations) of the spatial standard deviation maps. 
Therefore, the solar wind prediction using all the MCRs for the 
three Carrington rotations selected takes many days of CPU time, 
achieved in a reasonable time-span with the help of a start-up 
allocation on Comet at the San Diego Supercomputer Center (SDSC) 
of the National Science Foundation's Extreme Science and Engineering 
Discovery Environment \citep[XSEDE:][]{xsede}.

%
%
%
%

To determine the uncertainties in the model predictions, we 
compared the simulated CH and NL locations with the observed 
coronal holes and streamer locations \citep{jin13,pet09}, and 
compared the predicted \sws$\;$with in situ measurements near 
1~AU. For this, we used the coronal synoptic maps derived from 
the full--disk EUV images in 195~\AA$\;$data from 
\textit{STEREO}/SECCHI telescopes and the multispacecraft 
compilation of solar wind data from the OMNIweb archives
(\url{http://omniweb.gsfc.nasa.gov}). All these data are 
publicly available. 

\subsection{SPATIAL STANDARD DEVIATION MAPS}
            \label{sub:variance}

We used the 12--minute averaged longitudinal magnetograms from 
the HMI m\_720s series \citep{sch12} and the fully disambiguated 
vector magnetograms from the b\_720s series \citep{hoe14}. The 
full--disk magnetograms are computed every 12 minutes (720 seconds) 
by combining registered filtergrams obtained over a 1260~seconds 
time interval by the Vector Field camera. The spatial resolution 
is 1~arcsecond (half arcsecond pixels) and the full--disk images 
are collected on a $4096\times 4096$ detector. For the longitudinal 
magnetograms the noise level is nominally between 5 and 10~Gauss.
  
HMI uses a modified version of the Very Fast Inversion of the 
Stokes Vector (VFISV) code originally developed by \citet{bor11} 
to infer the vector magnetic field of the solar photosphere from 
its Stokes measurements. The 180--degree disambiguation in the 
direction of the magnetic field component transverse to the line 
of sight is addressed by HMI using different approaches. For 
strong--field regions, a variant of the {\em Minimum Energy} 
method proposed by \citet{met94} is implemented. In weak--field 
regions -- dominated by noise -- HMI provides results from 
three different methods: 
\begin{inparaenum} 
  \item[Method~1] selects the azimuth that is most closely aligned 
        with the potential field whose derivative is used in 
        approximating the gradient of the field;  
  \item[Method~2] assigns a random disambiguation for the azimuth;
  \item[Method~3] selects the azimuth that results in the field 
        vector being closest to radial. 
\end{inparaenum}
Methods~1 and 3 include information from the inversion, but can 
produce large--scale patterns in azimuth. Method~2 does not take 
advantage of any information available from the polarization, but 
does not exhibit any large--scale patterns and reflects the true 
uncertainty when the inversion returns purely noise. We, in this 
work, used the results from the {\em random disambiguation method}.

All HMI magnetograms have been processed through the well 
established NSO SOLIS/VSM pipeline. This pipeline already provides 
similar products derived from VSM observations.  Two different data 
products are generated for all Carrington rotations during solar 
cycle~24 covered by HMI since 2010: 
\begin{inparaenum} 
  \item[(1)] Integral Carrington synoptic maps of the magnetic 
        flux density for a selected number of Carrington rotations; 
  \item[(2)] Estimated spatial standard deviation maps associated 
        with each of those magnetic flux density maps. 
\end{inparaenum}
The necessary steps to generate these products are described in 
\citet{ber14}. To summarize, the procedure identifies the pixels 
in a set of full--disk magnetograms that contribute to a given 
heliographic bin in the synoptic map. Their number can vary quite 
significantly depending on the spatial resolution of the full--disk 
magnetograms, the size of the set, and location in latitude of the 
bin. Typically, lower latitudinal bins will contain a much larger 
number of pixels than those located in the solar polar regions. The 
weighted average and standard deviation are then computed from the 
magnetic field values of those pixels, resulting in two maps such 
as those shown in Figure~\ref{fig:hmi_maps}. All maps are generated 
on a regular $360 \times 180$ grid in Carrington longitude--sine 
latitude coordinates -- a spatial resolution sufficient for driving 
the PFSS and CSSS models. 

The spatial standard deviation is a measure of the statistical 
dispersion of all the pixel values contributing to a particular 
heliographic bin in the synoptic maps. It is not an estimate of 
error in the magnetic field observations. The number of pixels 
from HMI full disk magnetograms (pixel size of 0.5~arcsec) that 
contribute to an individual heliographic bin in a 360$\times$180 
synoptic map is quite large, greater than 6,400 at the equator. 
This number decreases for bins located at higher latitude bands 
but is significantly higher than, for example, the case of 
SOLIS/VSM observations discussed in \citep{ber14}. 

It should be noted that the spatial variance is expected to be 
higher in areas associated with strong fields, as compared to 
regions of quiet Sun. However, there is a significant difference 
between the results derived from SOLIS/VSM longitudinal magnetic 
field observations \citep[discussed in ][]{ber14} and those from 
HMI observations. The HMI spatial standard deviation maps show 
much higher values in areas of quiet Sun, up to about 100~Gauss 
near the polar regions. For comparison, this number is around 
10~Gauss for SOLIS/VSM (see Figure~4 in \citep{ber14}. This is 
due to the higher noise level in the HMI vector magnetic field 
measurements, particularly in the transverse component, compared 
to the SOLIS/VSM longitudinal measurements. This quiet Sun bias 
is removed from the final map in order to avoid overestimating 
the contribution of active patches to the standard deviation. The 
bottom image in Figure~2 reflects this correction.

Only two observations a day, taken at 0~UT and 12~UT, are used 
to construct the Carrington maps. This selection corresponds to 
times of high line--of--sight orbital velocity, that could affect 
the noise level in the magnetograms for some applications. However, 
a check using observations taken at 6:30~UT and 18:30~UT, around 
minimum relative orbital speed, has shown no significant 
differences in the final maps. In addition, we have also verified 
that using all HMI observations taken during a full Carrington 
rotation has no significant impact in the resulting low--resolution 
maps. Poorly observed polar regions are filled in using a 
cubic--polynomial surface fit to the currently observed fields 
at neighboring latitudes.  The fit is performed on a 
polar--projection of the map using low standard-deviation-to-fit 
measurements only, and the high--latitude fit is then integrated 
into the observed synoptic map, weighting toward the pole.

To test how the uncertainties in a synoptic magnetic flux density 
map may affect the calculation of the global magnetic field of 
the solar corona and the predicted solar wind speed at the source 
surface, we produced an ensemble of synoptic magnetic flux density 
maps generated from the standard deviation map for the three 
Carrington rotation presented in this paper. In the simulated 
synoptic maps of the ensemble, the value of each bin is randomly 
computed from a normal distribution with a mean equal to the 
magnetic flux value of the original bin and a standard deviation 
of $\sigma$, with $\sigma$ being the value of the corresponding 
bin in the standard deviation map. Figure~\ref{fig:hmi_maps} shows 
an example of HMI vector/longitudinal photospheric magnetic flux 
density synoptic map and the corresponding spatial standard 
deviation map (available at
\url{http://solis.nso.edu/0/vsm/vsm_maps.php}). The charts were 
computed using the radial component of the available inverted 
HMI fully--disambiguated, full--disk magnetograms covering 
Carrington rotation CR~2102. Due to the relatively high noise per 
pixel in the HMI measurements, the computed standard deviation 
map shows in general quite large values in regions of quiet Sun. 
These values increase quadratically from about 30~Gauss near 
the equator to about 100~Gauss in the polar regions. Since the 
quiet Sun has no effect on the outcome of the model, this 
dependency was removed from the map. The standard deviation map 
exhibits several properties. First, the values are significantly 
larger in areas associated with strong fields of active regions. 
This can be related to higher degree of spatial variation in 
magnetic structure as compared with quiet sun areas, and time 
evolution. Second, most areas with more uniform fields 
(\textit {e.~g}, coronal holes) show the smallest variance.     

One caveat is that the processing of HMI synoptic maps through 
the NSO SOLIS/VSM pipeline is not intended to solve any of the 
long--standing problem of the present--day synoptic maps such 
as the lack of information on the far--side and polar magnetic 
fields, and the open--flux problem \citep{lin17}. This is because 
we cannot design an HMI magnetogram for this purpose without a 
major recalibration based on cross--calibration of magnetograms 
from different observatories. The purpose of the present 
work in this paper is to test the new HMI mean--magnetic and 
spatial--variance maps (see also ``Synoptic Maps based on 
{\textit SDO}/HMI observations" at 
\url{https://solis.nso.edu/0/vsm/vsm\_maps.php}) and determine 
the uncertainties in the solar wind prediction. 

\subsection{METRICS OF ACCURACY}  
            \label{sub:metrics}
               
Comparing the correlation coefficient between the observed (OMNI 
data) and predicted \sws$\;$alone may be insufficient to assess 
the predictive capabilities of the coronal models since 
correlation coefficients do not capture scaling differences 
between the observed and predicted quantities as pointed out 
in \citet{pod14} and \citet{pod16}. Therefore, we obtained the 
RMSEs between the observed \sws$\;$and the speeds predicted by 
the CSSS model. 

%
%
\section{METHOD}
         \label{sec:method}

The uncertainty estimate of solar wind prediction presented in 
this paper is based primarily on the computations of the CSSS model. 
We obtained the global coronal magnetic field by extrapolating the observed 
photospheric magnetic field (the ensemble of HMI magnetic flux 
density synoptic maps) to the corona (and beyond) using the 
CSSS model and subsequently, the FTEs according to 
Equation~\ref{eq:fte} and predicted \sws$\;$at 2.5~\rsun$\;$using 
the empirical relationship shown in Table~\ref{tab:speedfte}. 
We then propagated the predicted \sws$\;$kinematically to 1~AU
to compare with the OMNI data of observed solar wind.. 

\begin{figure}
  \centerline{\hspace*{-0.025\textwidth}
  \includegraphics[width=0.515\textwidth,clip=]{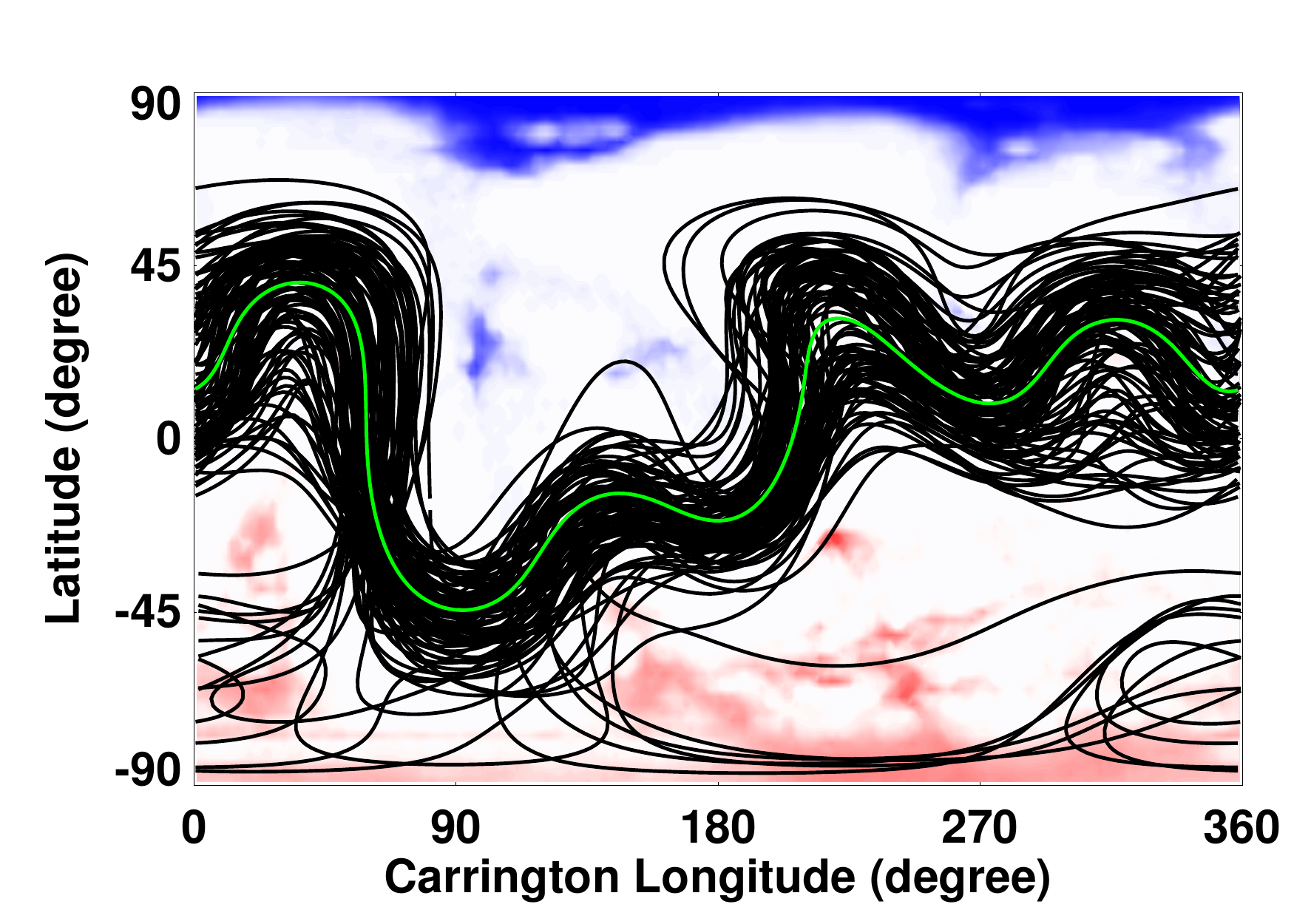}
  \hspace*{-0.03\textwidth}
  \includegraphics[width=0.515\textwidth,clip=]{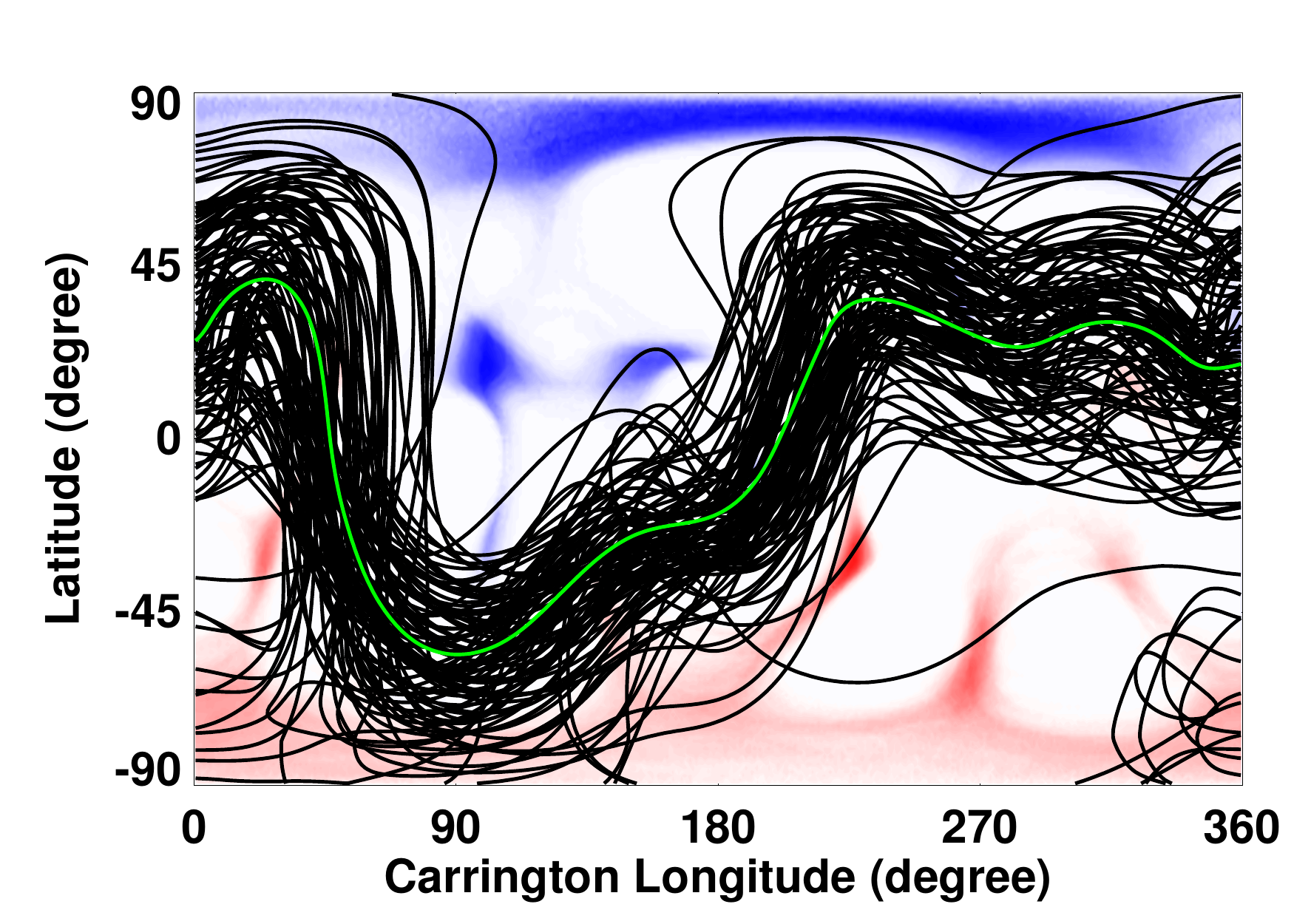}
          }         
  \vspace{-0.38\textwidth} 
  \centerline{\small \bf    
   \hspace{0.08 \textwidth} \color{black}{(a) PFSS model: CR 2102}
  \hspace{0.13\textwidth}  \color{black}{(b) CSSS model: CR 2102}
          \hfill}
  \vspace{0.34\textwidth}  
  \centerline{\hspace*{-0.025\textwidth}
  \includegraphics[width=0.515\textwidth,clip=]{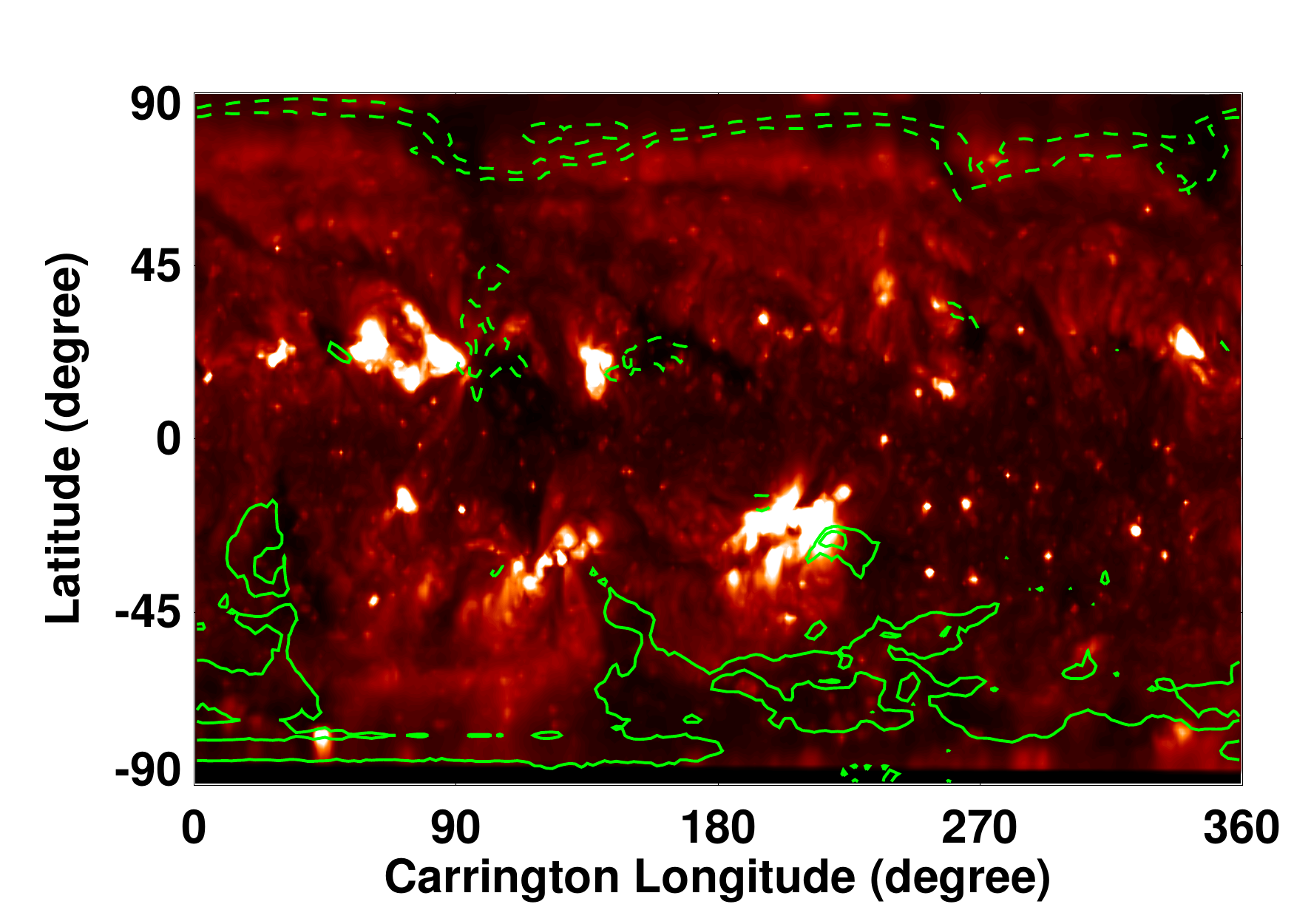}
  \hspace*{-0.03\textwidth}
  \includegraphics[width=0.515\textwidth,clip=]{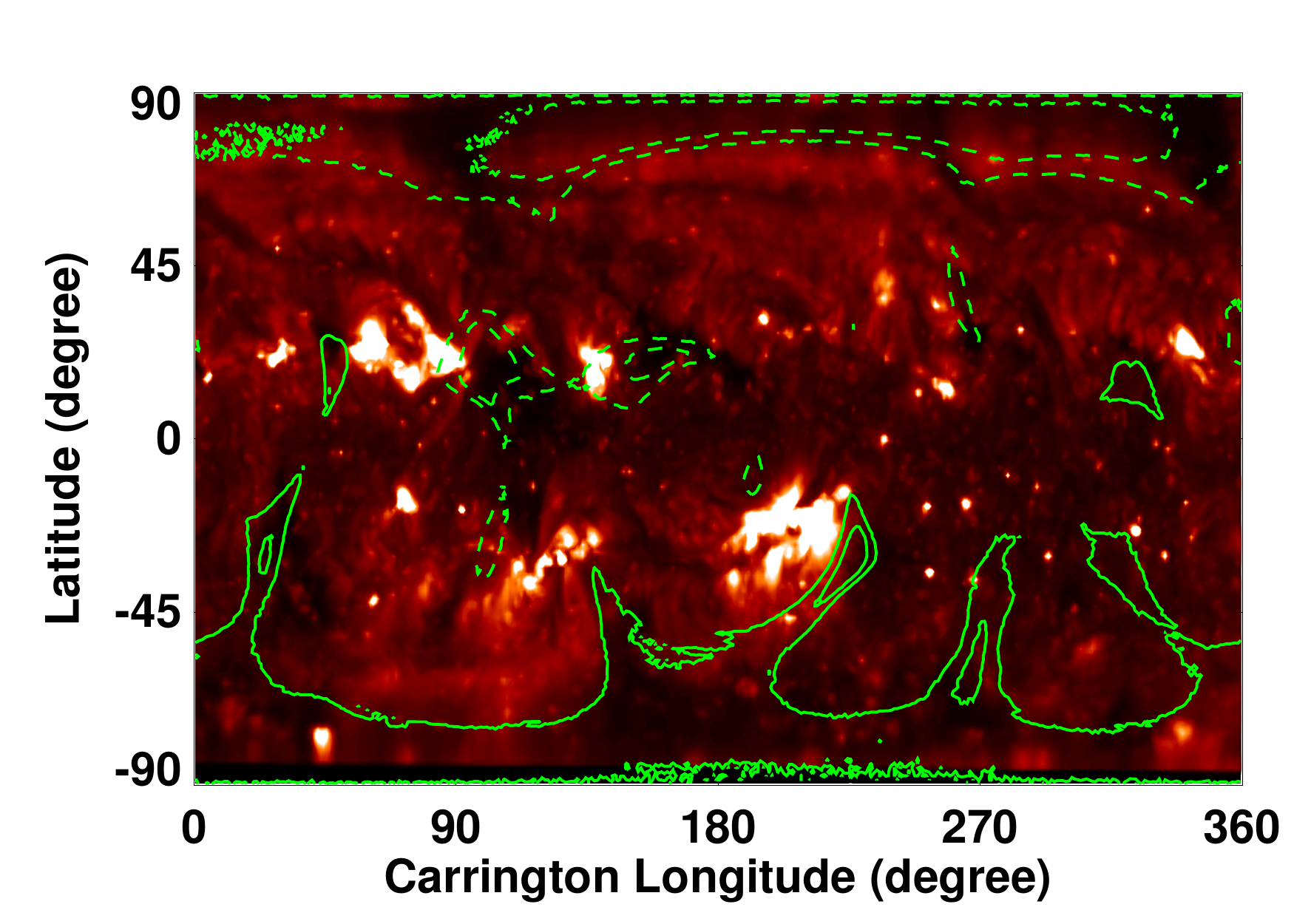}
          } 
  \vspace{-0.38\textwidth} 
  \centerline{\small \bf   
  \hspace{0.08 \textwidth} \color{black}{(c) PFSS model: CR 2102 }
  \hspace{0.13\textwidth}  \color{black}{(d) CSSS model: CR 2102}
        \hfill}
  \vspace{0.345\textwidth} 
  \caption{Panels~(a) and (b): The solid black lines represent 
           the NLs obtained for Monte-Carlo simulations of 
           magnetic flux synoptic maps generated from the standard 
           deviation map for: PFSS (left) and CSSS (right) models 
           for CR~2102. The NLs associated with the original 
           synoptic map are over--plotted in green for both the 
           models. Positive/negative open field footpoints 
           (red/blue pixels) are also shown in a color scale 
           with stronger/fainter coloring indicating where a 
           larger/smaller fraction of the models have open fields. 
           Panels~(c) and (d): the intensity distribution at 
           wavelength 195\AA$\;$from \textit {STEREO}/SECCHI for 
           CR~2102. Over--plotted in green are contours of the 
           PFSS (left) and CSSS (right) CH distributions shown in 
           the top panels. Solid/dashed lines represent 
           positive/negative CH contours. The image has been 
           enhanced to show the locations of coronal holes (dark 
           regions).
          }
          \label{fig:chnl1} 
\end{figure}

\begin{figure}
   \centerline{\hspace*{-0.025\textwidth}
   \includegraphics[width=0.515\textwidth,clip=]{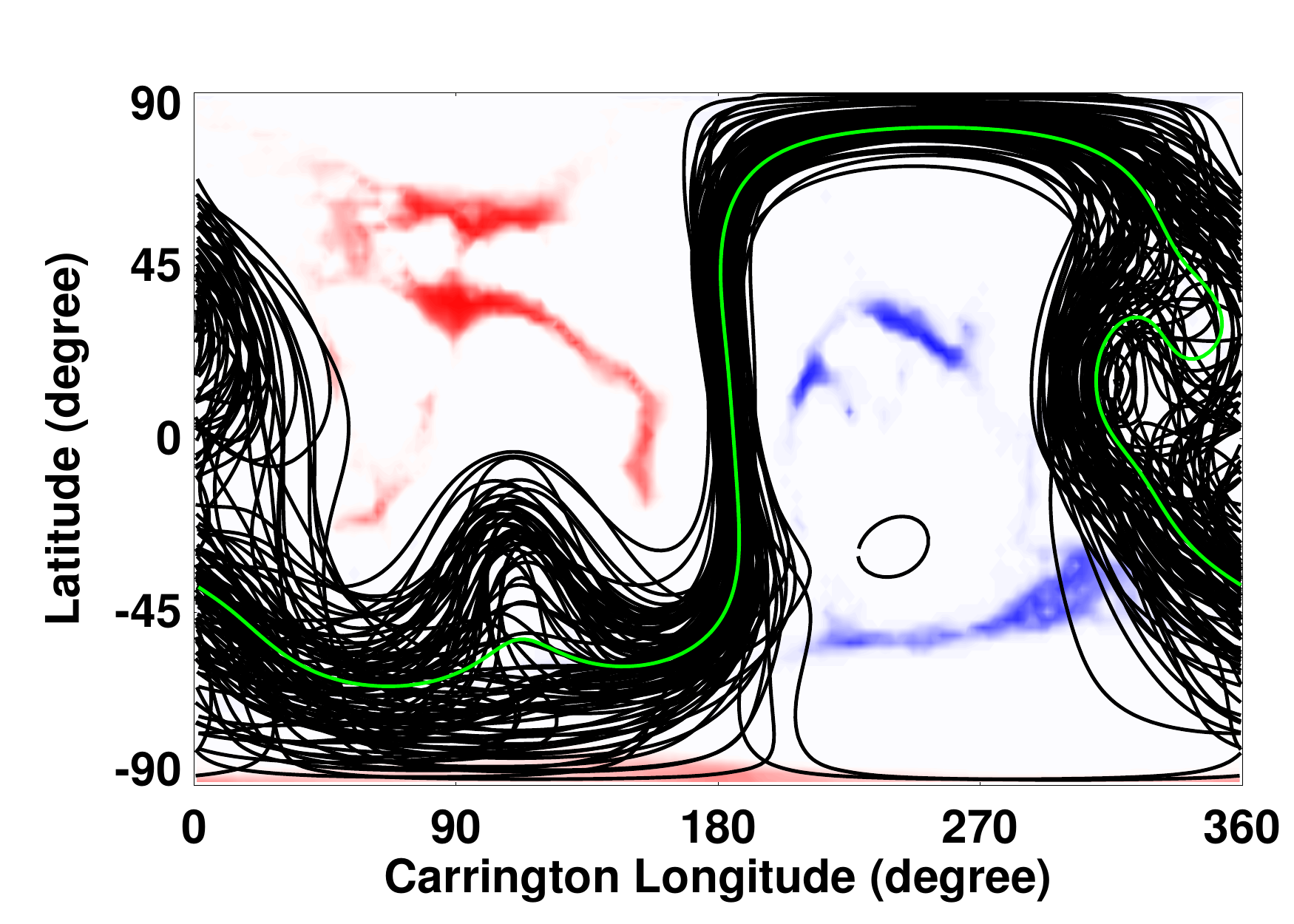}
   \hspace*{-0.03\textwidth}
   \includegraphics[width=0.515\textwidth,clip=]{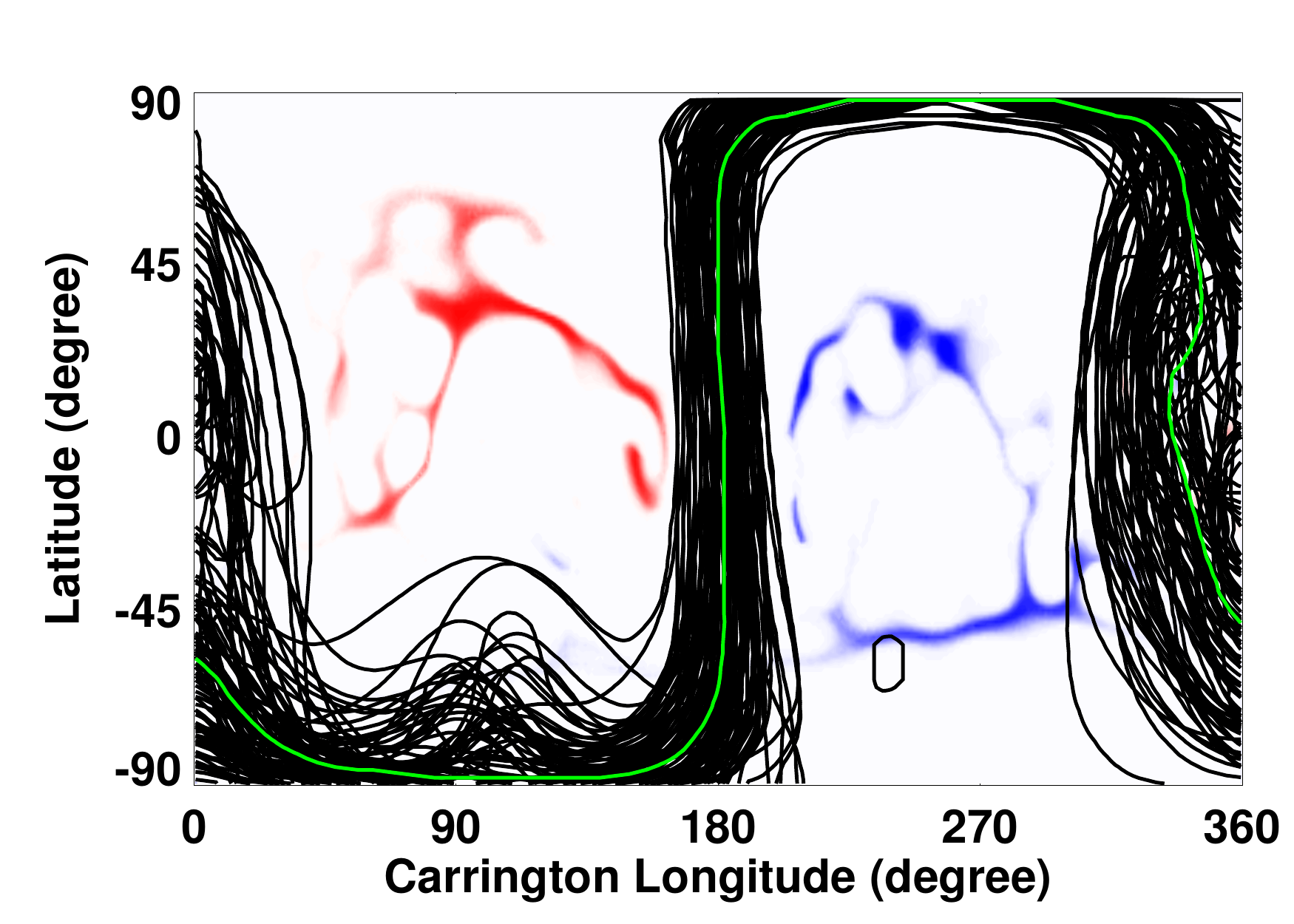}}
   \vspace{-0.38\textwidth} 
   \centerline{\small \bf    
   \hspace{0.08 \textwidth} \color{black}{(a) PFSS model: CR 2137}
   \hspace{0.13\textwidth}  \color{black}{(b) CSSS model: CR 2137}
         \hfill}
   \vspace{0.34\textwidth}  
   \centerline{\hspace*{-0.025\textwidth}
   \includegraphics[width=0.515\textwidth,clip=]{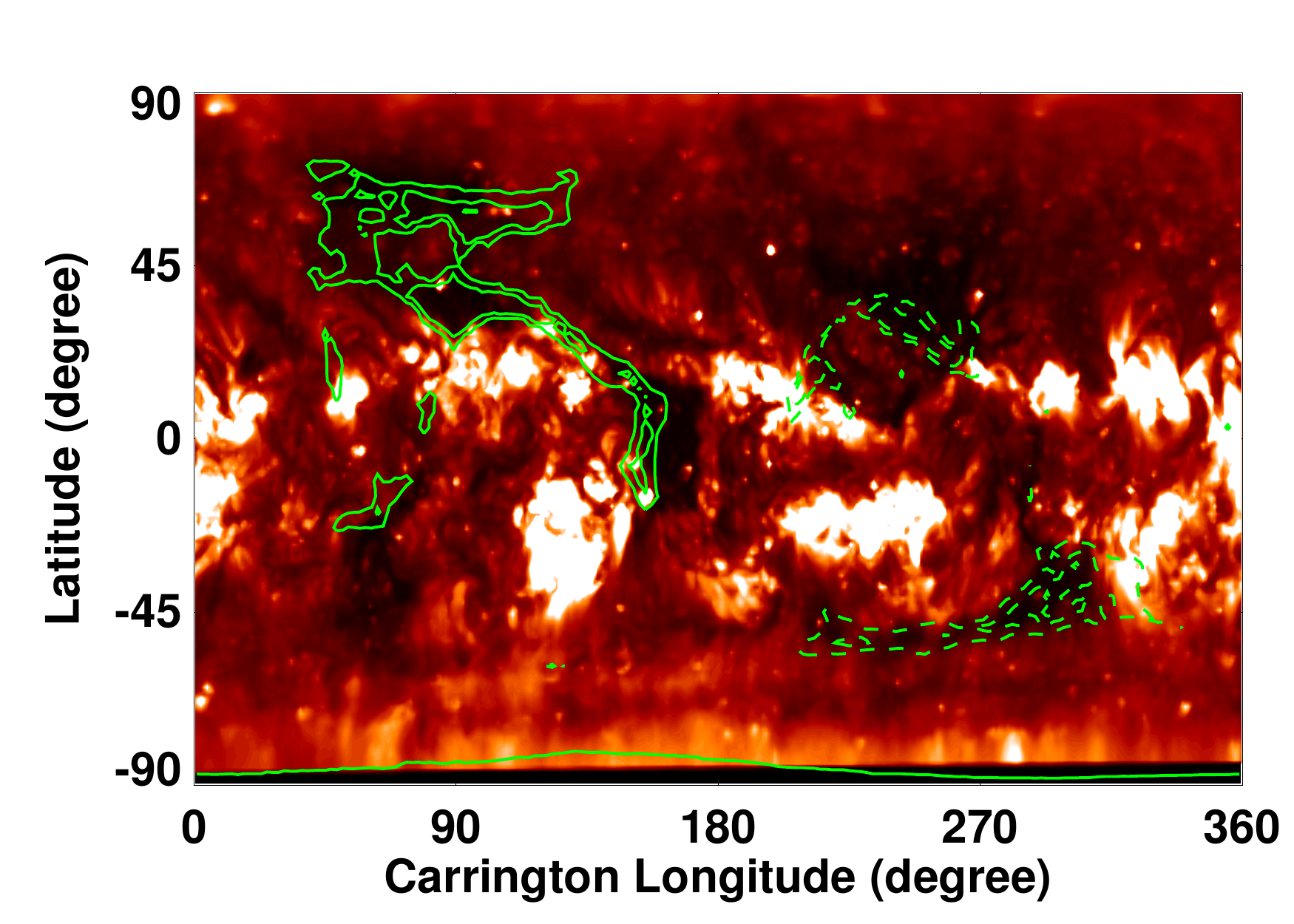}
   \hspace*{-0.03\textwidth}
   \includegraphics[width=0.515\textwidth,clip=]{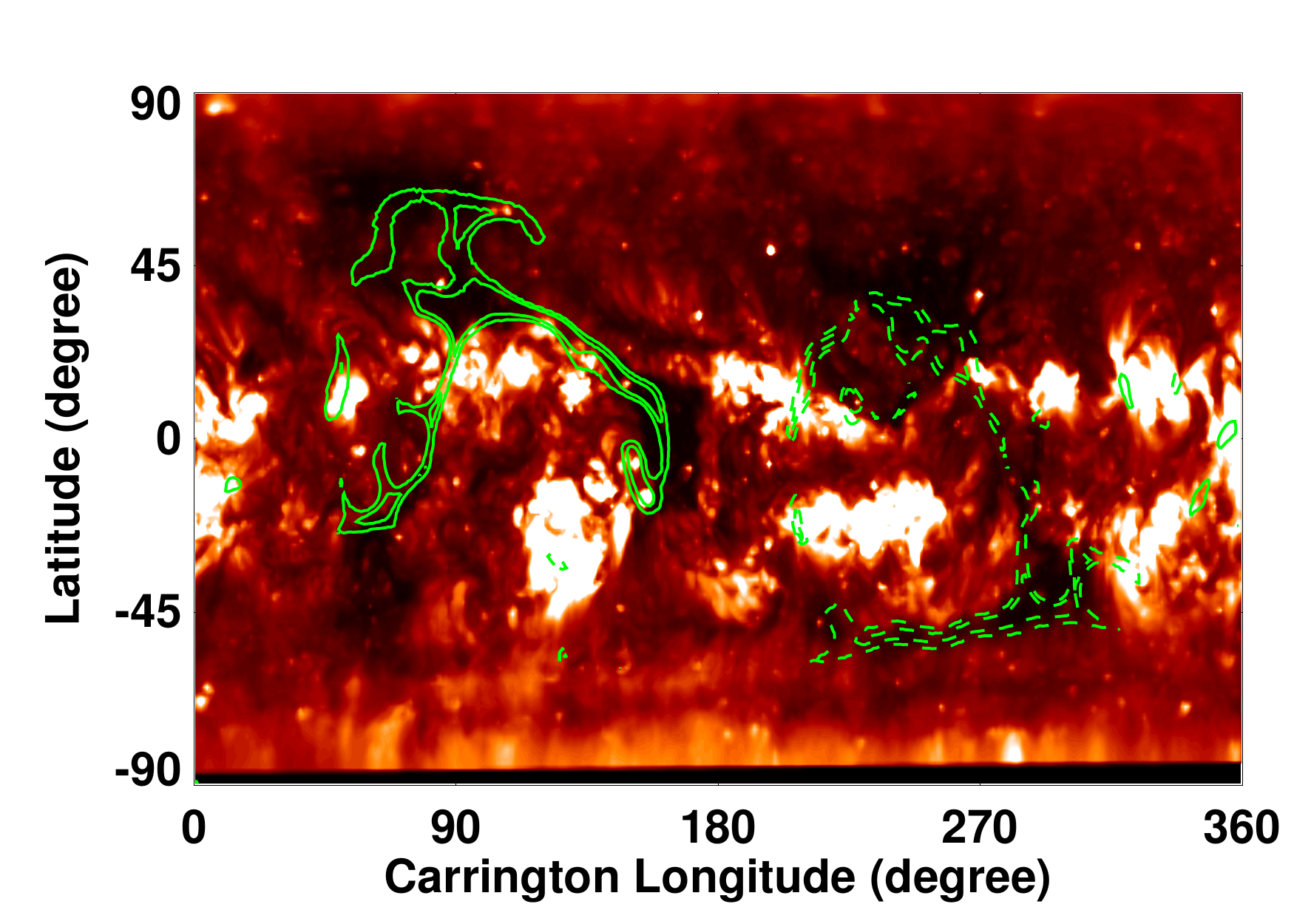}}
   \vspace{-0.38\textwidth} 
   \centerline{\small \bf  
   \hspace{0.08 \textwidth} \color{black}{(c) PFSS model: CR 2137}
   \hspace{0.13\textwidth}  \color{black}{(d) CSSS model: CR 2137}
   \hfill}
   \vspace{0.345\textwidth}  
   \caption{Same as Figure~\ref{fig:chnl1} but for CR~2137. }
          \label{fig:chnl2} 
\end{figure}

\begin{figure}
  \centerline{\hspace*{-0.025\textwidth}
  \includegraphics[width=0.515\textwidth,clip=]{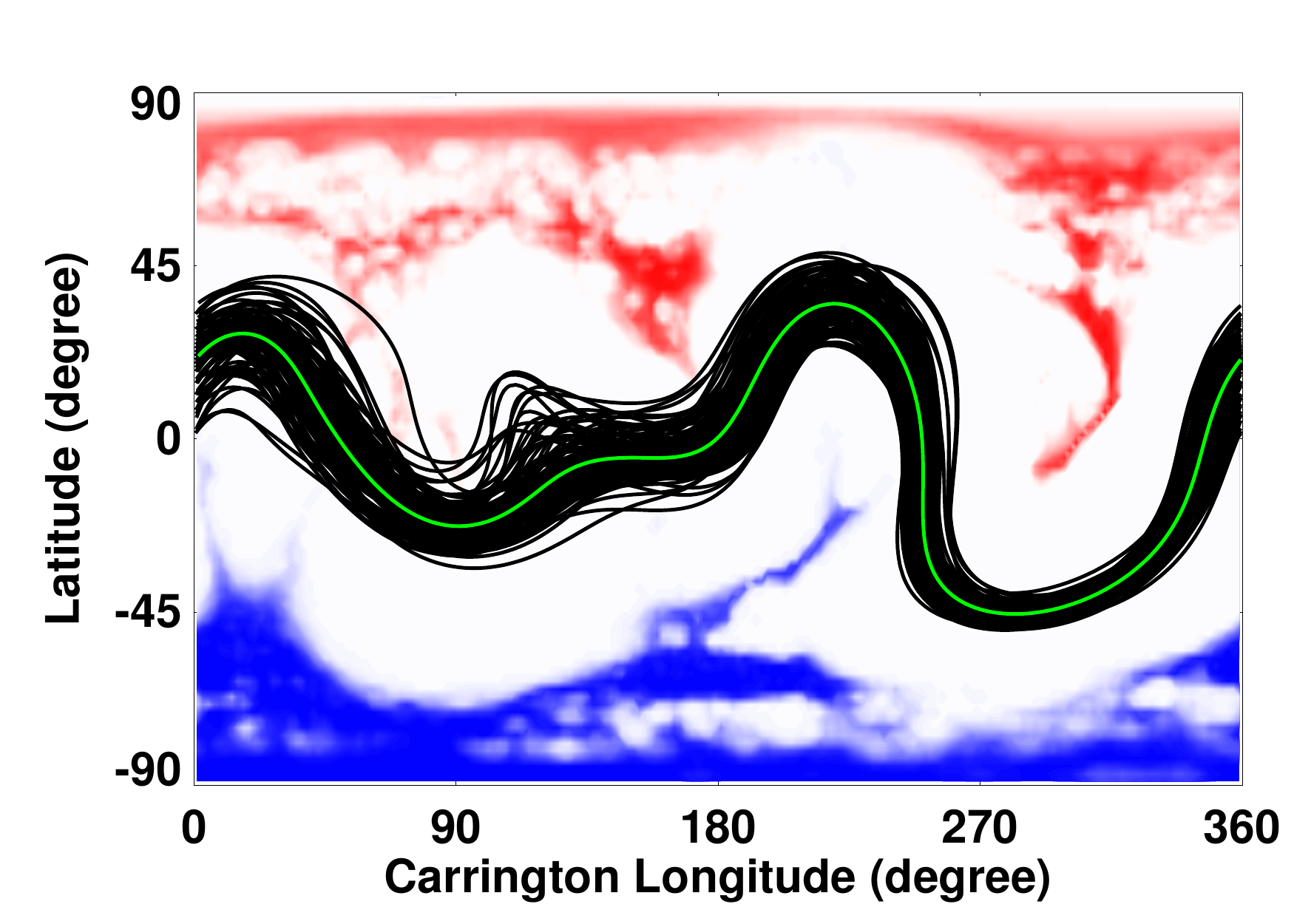}
  \hspace*{-0.03\textwidth}
  \includegraphics[width=0.515\textwidth,clip=]{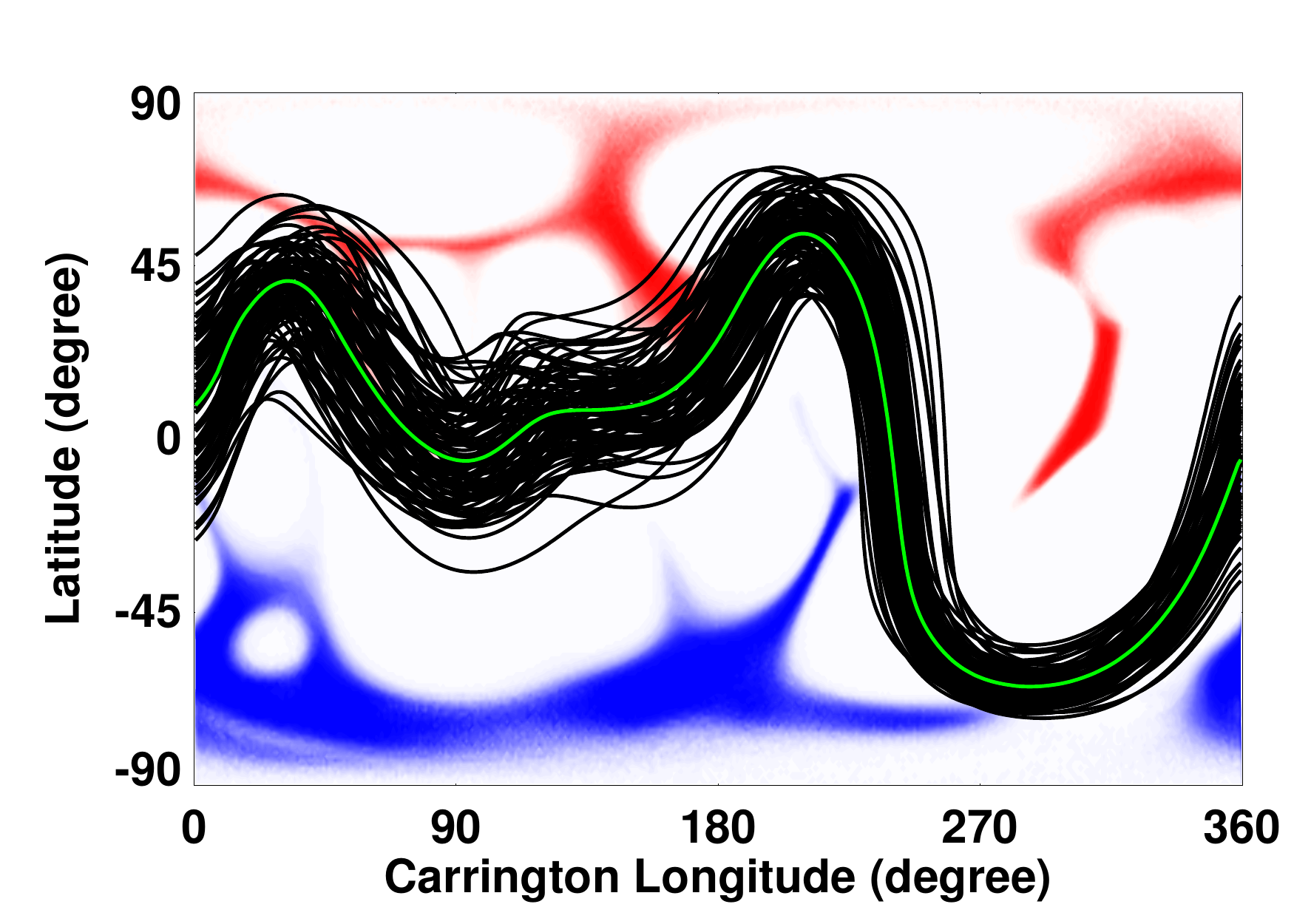}}
  \vspace{-0.38\textwidth}
  \centerline{\small \bf 
  \hspace{0.07 \textwidth} \color{black}{(a) PFSS model: CR 2160}
  \hspace{0.13\textwidth}  \color{black}{(b) CSSS model: CR 2160}
  \hfill}
  \vspace{0.34\textwidth}
  \centerline{\hspace*{-0.025\textwidth}
  \includegraphics[width=0.515\textwidth,clip=]{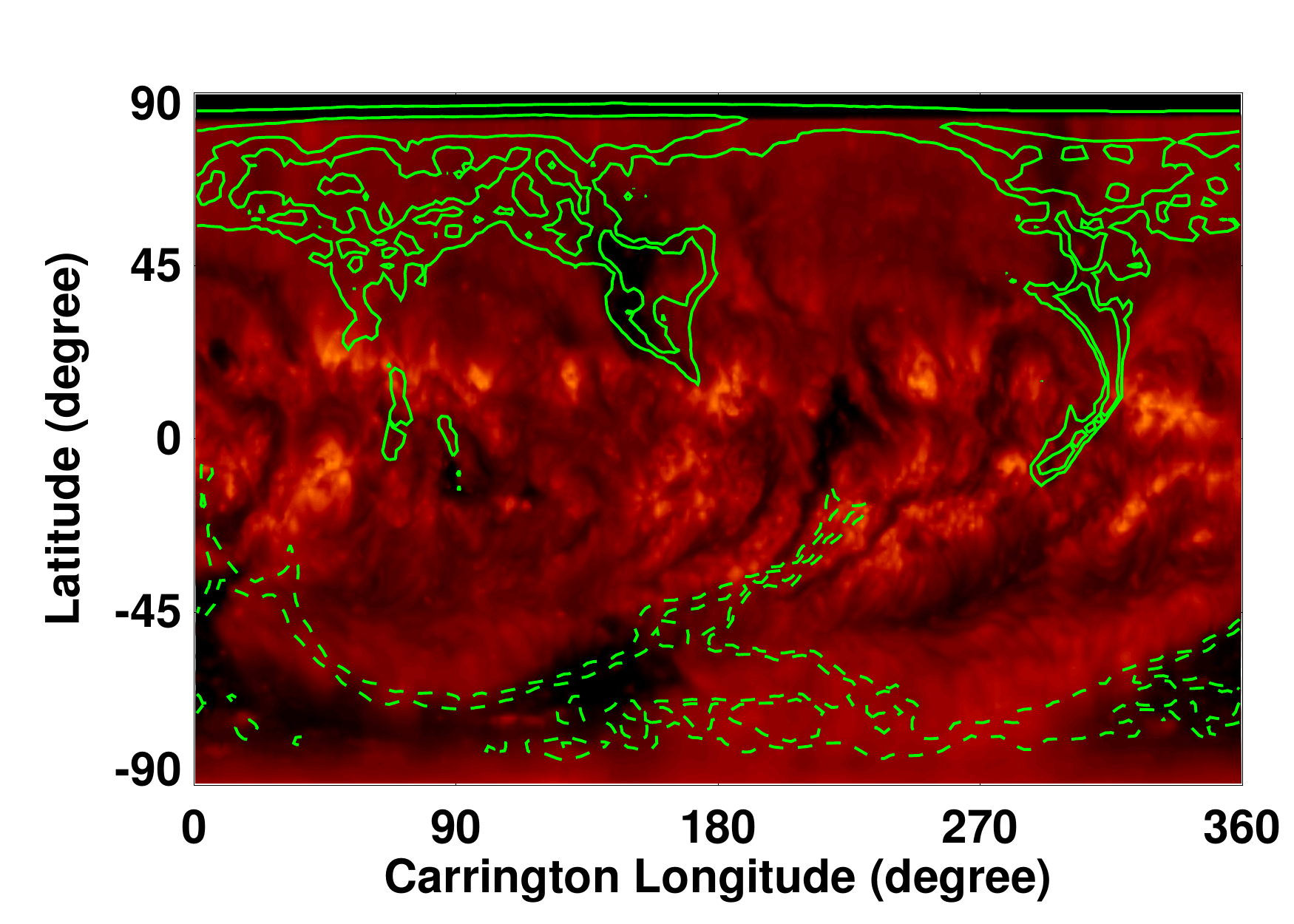}
  \hspace*{-0.03\textwidth}
  \includegraphics[width=0.515\textwidth,clip=]{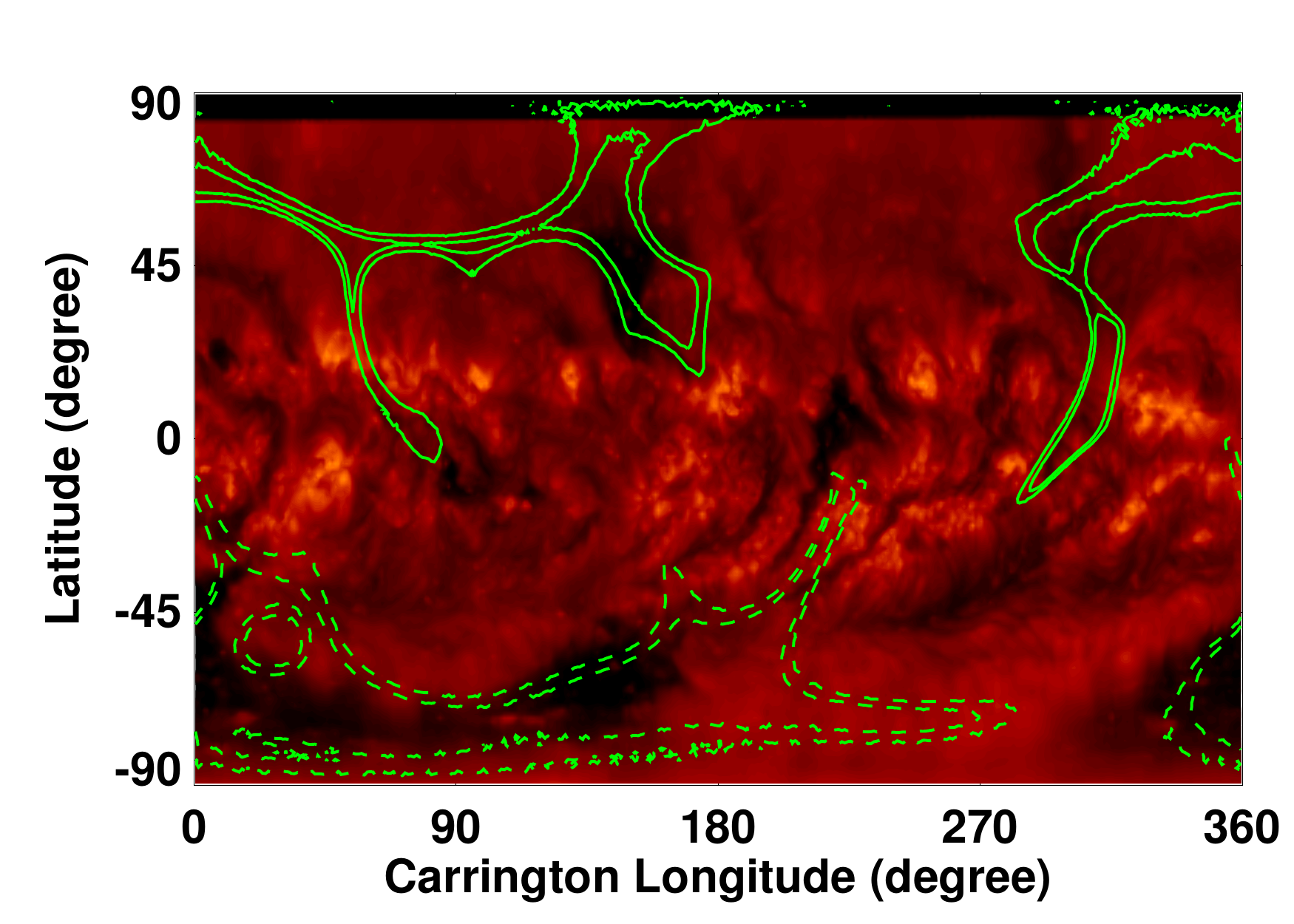}}
  \vspace{-0.38\textwidth} 
  \centerline{\small \bf  
  \hspace{0.07 \textwidth} \color{black}{(c) PFSS model: CR 2160}
  \hspace{0.13\textwidth}  \color{black}{(d) CSSS model: CR 2160}
  \hfill}
  \vspace{0.345\textwidth}
  \caption{Same as Figure~\ref{fig:chnl1} but for CR 2160.}
          \label{fig:chnl3} 
\end{figure}

We computed the footpoint locations of the open field lines 
(CHs) and the magnetic neutral lines (NLs) 
(Figures~\ref{fig:chnl1} -- \ref{fig:chnl3}) using the CSSS 
model and compared with those computed using the widely used 
PFSS model. We utilized the numerical elliptic solver available 
in MUDPACK of National Center for Atmospheric Research 
(\url{http://www2.cisl.ucar.edu/resources/legacy/mudpack)}) 
finite--difference package: \citep[see][for details]{pet13} 
for the PFSS computations. Such a comparison will provide 
additional validation of the predictive capability of the CSSS 
model \citep[see][for CSSS model validation.]{pod14,pod16}. In 
all the model computations, we used the ``HMI synoptic maps"
and the corresponding ``MCRs" as the lower boundary conditions 
for each of the three Carrington rotations selected for the study. 

While a complete 3-dimensional MHD model may represent the corona 
more realistically, magnetostatic models such as the PFSS and the 
CSSS models are computationally inexpensive and much faster. Radial 
lower boundary data from photospheric vector measurements are 
appropriate for this problem because they give us genuine observations 
of the radial flux distribution, which is not the case with the 
longitudinal field measurements usually employed in global 
coronal/heliospheric modeling.

For the solar wind prediction, we adopted a two--step method 
similar to \citet{pod14} and \citet{pod16} with the exception 
that instead of mapping the observed solar wind back to the 
corona, we computed the \sws$\;$at 2.5~\rsun$\;$and propagated 
kinematically to 1~AU for comparison with in situ observations.  

In Step~1, we computed FTEs on a Carrington 
longitude--heliographic latitude grid of $1^\circ$ at 
2.5~\rsun$\;$corresponding to the cusp surface in the CSSS model 
and ``predicted'' \sws$\;$by making use of the WS empirical 
relationship shown in Table~\ref{tab:speedfte} 
\citep{wan90a,wan95,wan97}. We computed FTE at the cusp surface 
(at 2.5~\rsun) because the magnetic flux tube expansion rate 
that influences the solar wind speed is relevant and significant 
at the base of the corona where the magnetic field lines begin 
to open, namely, the cusp surface \citep{pod16}. The field lines 
become radial at the source surface at 15~\rsun. Then we obtained 
a quadratic function \citep[Figure~2 in ][]{pod14} that best 
fitted the pair, predicted--speed/computed--FTE. We adopted this 
method because the quadratic equation best represents the 
nonlinear, super--radial expansion of the magnetic flux tubes 
and is physically intuitive, though simple. This approach is 
different from that of \citet{arg00}, modified in \citet{mcg11b}. 
Moreover, as shown in \citep{pod16}, the temporal variation of 
the coefficients provides a strong implication of the influence 
of the changing magnetic field conditions on the solar wind 
outflow. 

In order to obtain the coefficients of the best--fit quadratic
equation and the subsequent solar wind speed prediction for the
three selected Carrington rotations, CRs~2102, 2137 and 2160, 
we used the original
NSO processed HMI synoptic maps over a four--Carrington rotation 
period that included the respective CRs chosen for the present 
study. Our Step~2 consisted of predicting solar wind speeds 
using the FTEs computed for each of~98 Monte-Carlo realizations 
of the standard deviation maps described earlier and the fitted 
quadratic equations to obtain the predicted \sws$\;$at 
2.5\rsun$\;$for each of the selected rotations. 

%
%
%
%
%
%
%
%
%
%
%
\begin{figure}[!ht]
  \vspace{-1.4in}
  \centerline{
    \hspace*{0.01\textwidth}
    \includegraphics[width=1.0\textwidth,clip=]{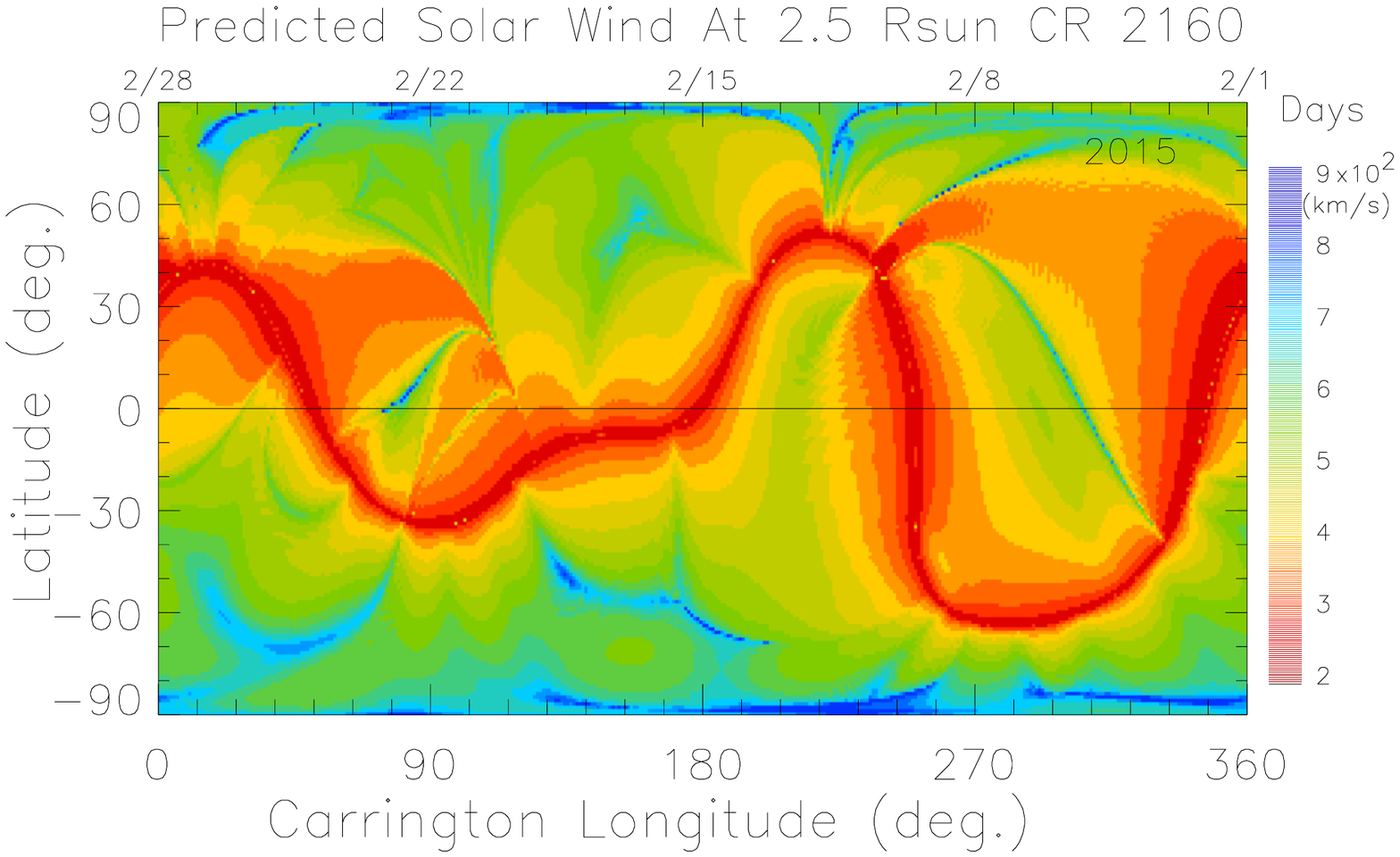}
    \hspace*{-0.9\textwidth}
    \includegraphics[width=0.79\textwidth]{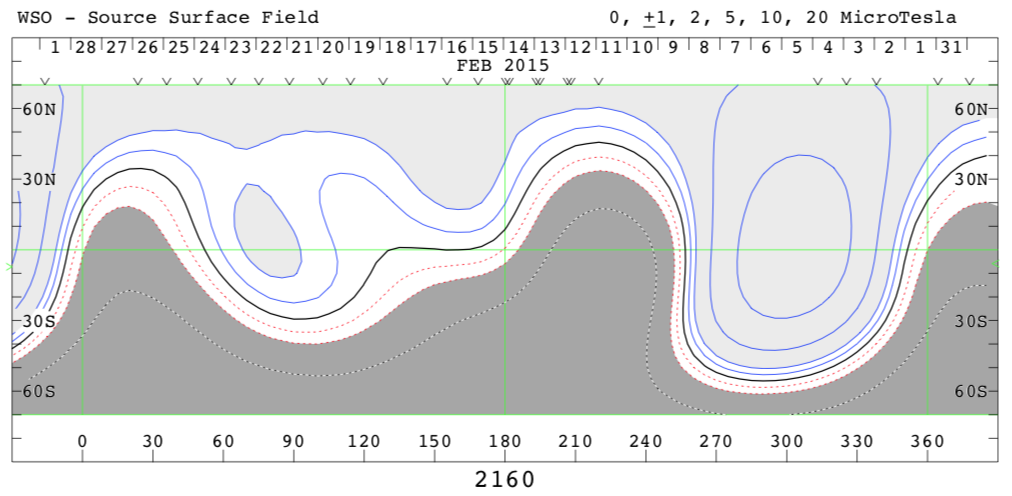}
  } 
  \caption{ {\bf Top panel:} The V-map (solar wind velocity synoptic 
            map) at 2.5\rsun$\;$created using the  solar wind speed 
            predicted using the CSSS model for the Carrington 
            Rotation CR~2160 (1--28   February 2015). The slow solar 
            wind (speed $\le 450$ \kms) is represented by the red 
            color while fast wind (speed $ > $ 750 \kms)  is shown 
            in blue color. The slow solar wind belt corresponds to 
            the heliospheric current  sheet (HCS). 
            {\bf Bottom panel:} The coronal magnetic field computed
            using the Wilcox Solar Observatory (WSO) synoptic maps 
            and the PFSS model. The black solid line represents the 
            magnetic neutral  line (heliospheric current sheet -- HCS).
            Courtsey: J. T. Hoeksema 
            (\url{http://wso.stanford.edu/synsourcel.html}).
          }
          \label{fig:vmap}
\end{figure}
%
%
%
For the forward propagation of predicted near--Sun solar wind, 
we adopted a simple, kinematic model described in \citet{arg00} 
allowing for interaction between neighboring fast and slow 
streams to a limited extent. Using this approach, a solar wind 
velocity synoptic map, \textit{V--map} (Figure~\ref{fig:vmap}), 
is created at the source surface with the predicted \sws$\;$as 
described in \S~\ref{sec:method}. Then the solar wind is allowed 
to propagate at constant radial speed, the value computed at each 
grid point at the source surface, for a distance of 1/8~AU. At 
this point, the velocities were recalculated to allow for the 
interaction between fast and slow winds according to:
\begin{equation}
  v_i = \sqrt  \frac{2}{\frac{1}{v_i^2} + \frac{1}{v_{i+1}^2}}  
                   \label{eq:interact}
\end{equation}
where, $v_i$ is the solar wind speed at the $i^{th}$ grid. The 
new velocities are now used for propagating the solar wind to 
2/8~AU, where the velocities are recalculated according to 
Equation~\ref{eq:interact}. This is continued until the solar 
wind reaches 1~AU. 

%
%
\begin{figure}[htbp] 
  \centering 
  \vspace{-0.1in}
  \includegraphics[width=0.9\textwidth]{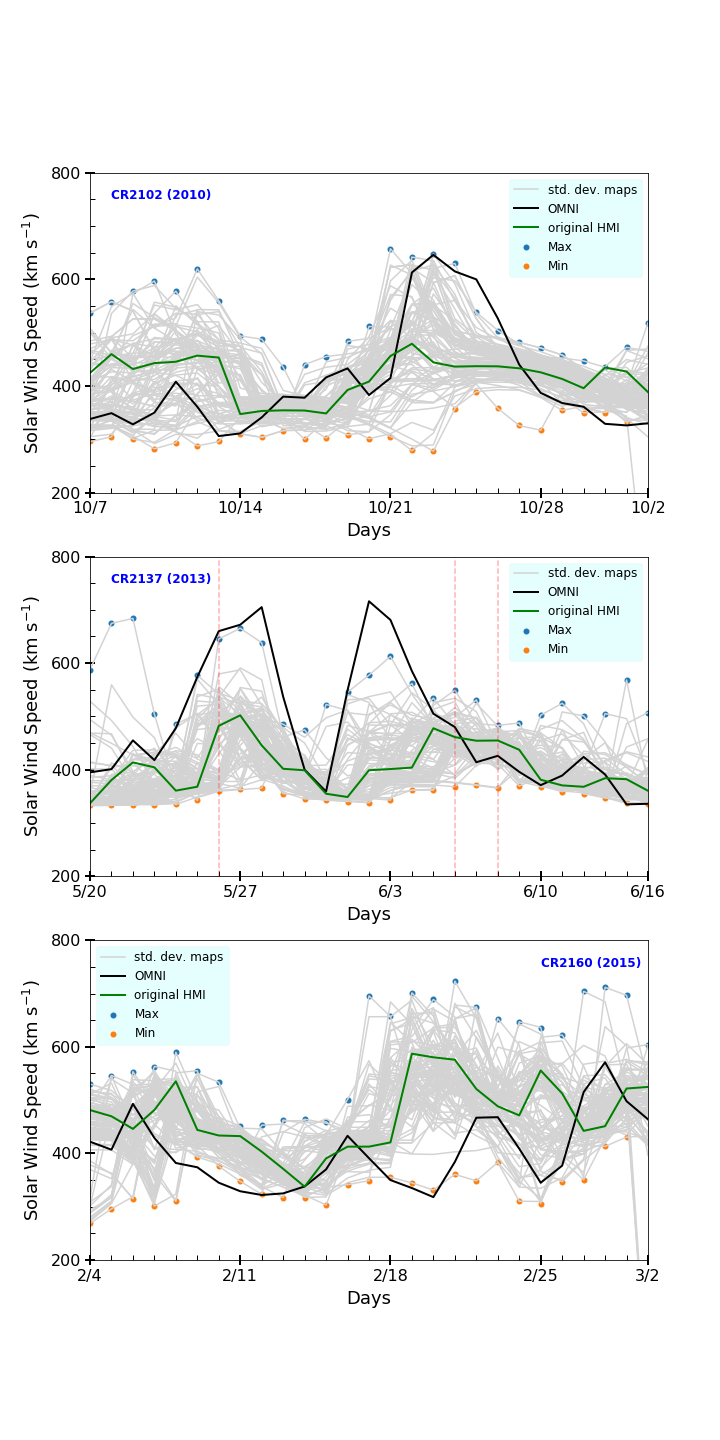}
  \vspace{-0.8in}
  \caption{Comparison: The gray solid lines represent the predicted 
           solar wind speed using the CSSS model 
           (\S~\ref{sec:method}) and the ensemble HMI spatial 
           standard deviation maps (\S~\ref{sub:variance})
           for Carrington rotations CRs~CR~2102 (3~--~30 October, 
           2010), 2137 (May~14 -- June~11, 2013) and 2160 (1~--~28 
           February, 2015), 2137. The black solid line depicts the 
           observed solar wind speed corresponding to the arrival 
           times of the predicted solar wind (see text for details) 
           and the green solid line represents the predicted speed 
           using the regular HMI synoptic map for the same rotation. 
           The blue and orange circles depict the maximum and 
           minimum envelopes in the ensemble predctions. The red 
           vertical lines in the middle panel (CR~2137 (2013)) 
           corresponding to May 26th and June 8th indicate the two 
           ICMEs listed in the Richardonson/Cane catalog 
           \citep{can03,ric10}. 
          }
          \label{fig:swspred}
\end{figure}
%

For obtaining the prediction accuracies, we used the daily 
averaged solar wind speed from the OMNI archive corresponding 
to the arrival times of the predicted solar wind. Since our 
model predictions are at a higher resolution, we took a 
$13^\circ$ average near the ecliptic corresponding to the 
observed $b0$ angle available in the OMNI data for better 
comparison. Figure~\ref{fig:swspred} shows the uncertainties 
(the spread) in the predicted solar wind speed for the 
98~MCR of the HMI spatial standard deviation 
maps for the three selected Carrington rotations. We, then, 
obtained the RMS errors between the CSSS predictions and the 
observed solar wind (Figure~\ref{fig:swserr}) for all the three 
Carrington rotations. 

%
%
\section{DISCUSSION OF THE RESULTS}
         \label{sec:results}

Presented in this paper is the influence of the uncertainties 
in the construction of photospheric magnetic flux density 
synoptic maps from the observed daily magnetograms on the 
coronal features and solar wind predicted using them. These
uncertainties are represented as the spatial standard deviation 
maps as shown in Figure~\ref{fig:hmi_maps}. As well known, the 
flux density synoptic maps serve as the inner boundary conditions 
for various coronal models \citep[e.g.][]{hoe84,ril15,lin17}, 
including the operational WSA model \citep{arg00}, for computing 
the coronal and interplanetary magnetic fields, coronal features 
and the solar wind speed. In this study, we obtained the predicted 
CH locations, NLs and \sws$\;$at 1~AU using the coronal 
extrapolation models, CSSS \& PFSS, and the spatial standard 
deviation synoptic maps \citep{ber14} derived from the 12--minute 
averaged full--disk \textit{SDO}/HMI longitudinal magnetograms 
(m\_720 series \citep{sch12}) and the fully disambiguated vector 
magnetograms (b\_720 series \citep{hoe14}) through the NSO 
SOLIS/VSM pipeline.

Figures~\ref{fig:chnl1} -- \ref{fig:chnl3} depict the locations 
of CHs and NLs computed using the ensemble of HMI spatial 
standard deviation maps for CR~2102, 2137 and 2160. The left 
columns show the results of PFSS model while the right columns 
represent the CSSS model. A visual inspection of the SECCHI 
195~\AA$\;$synoptic map reveals that the locations of CHs match 
reasonably well with the predicted CH locations. Also, the CH 
locations and the NLs predicted by the CSSS model matches with 
the predictions of PFSS model in general. However, we note that 
the northern high--latitude CH around ($90^{\circ}, 60^{\circ}$) 
is larger in the PFSS than the CSSS model, whereas the 
low--latitude CH structure around ($80^{\circ}, 0^{\circ}$) is 
more extensive in the CSSS than the PFSS model. These general 
agreement between the models and, between the model and 
observations validates the robustness of the CSSS model in 
predicting the coronal features and the \sws. Moreover, we note 
that there is a large spread in the computed neutral lines and 
a few are too far outside of the general trend of the neutral 
lines. Interestingly, the spread in the NLs is minimum for CR~2160 
which is close to the maximum phase of the solar cycle, contrary 
to the general expectations. The NLs and CHs are key features of 
the coronal models and the spread in the NL and the CH locations 
as seen in Figures~\ref{fig:chnl1} -- \ref{fig:chnl3} indicate 
the influence of the uncertainties in the photospheric magnetic 
field measurements on these features. 

The top panel of Figure~\ref{fig:vmap} depicts the synoptic map 
of the predicted solar wind speed, V-map, at 2.5~\rsun$\;$for 
CR~2160 (1~--~28 February, 2015). Here, the slow solar wind 
(speed $\le 450$ \kms) is depicted by red and the fast wind 
(speed $ > 750$ \kms) is represented by dark~blue. The bottom 
panel shows the coronal magnetic fields modeled by the PFSS model 
using the Wilcox Solar Observatory synoptic maps (Courtesy: J.~T. 
Hoeksema, \url{http://wso.stanford.edu/synsourcel.html}. The black 
solid line represents the heliospheric current sheet (HCS). The 
slow wind belt of the predicted solar wind speed generally follows 
the HCS as evidenced by a visual inspection of the two figures. 
Since the WSO synoptic maps have been extensively used as a 
standard for validating model predictions of global coronal 
structures for decades, this comparison provides further 
validation of the CSSS model in the prediction of solar wind.

%
%
\begin{figure}
  \centering
  \includegraphics[width=1.0\textwidth]{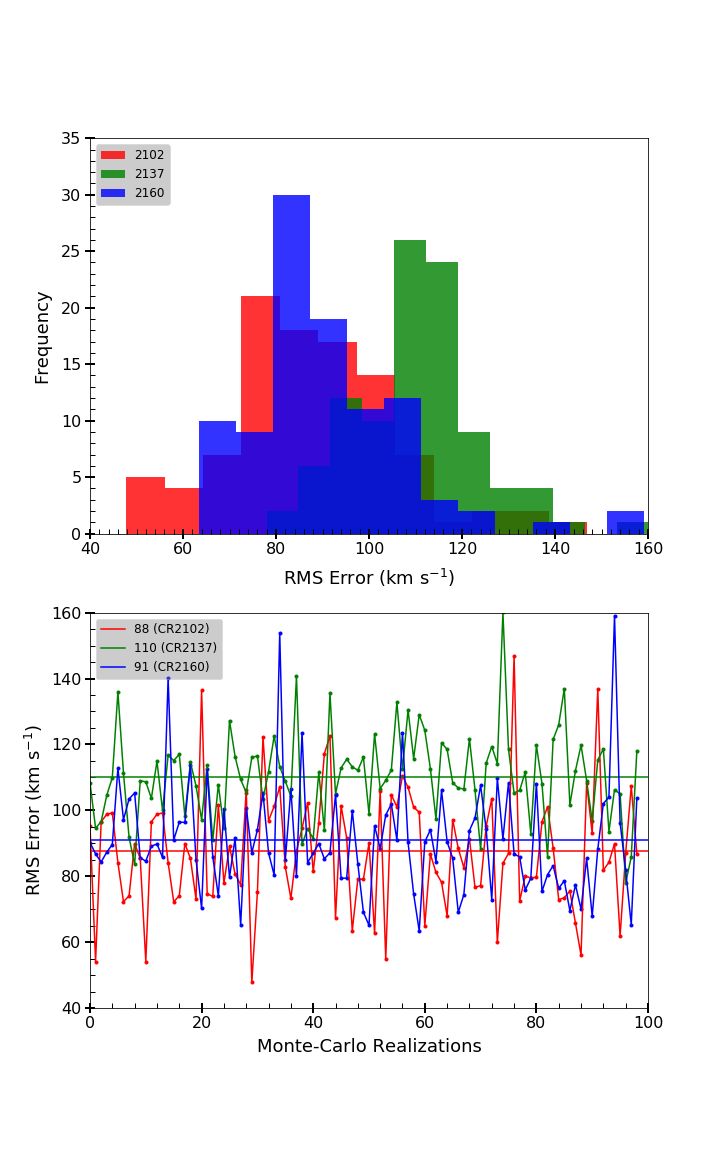}
  \vspace{-0.2in}
  \caption{The RMS errors between the solar wind speed 
           predicted using the CSSS model and that observed
           at 1~AU (OMNI data) for CRs~2102, 2137 and 2160 
           The mean RMSEs are 85, 110 and 92 respectively
           for the chosen Carrington rotations.
          }
          \label{fig:swserr}
  \vspace{-0.4in}
\end{figure}
%
%

Figure~\ref{fig:swspred} depicts a comparison of \sws$\;$predicted by CSSS 
model (gray solid lines) for each of the 98 spatial standard deviation maps for 
all the three Carrington rotations, CR~2102 (3~--~30 October, 2010), 
2137 (May~14 -- June~11, 2013) and 2160 (1~--~28 February, 2015), 
with the corresponding observed (OMNI) solar wind speed (black solid 
curve). The green line depicts the CSSS prediction for the regular 
HMI synoptic maps. The blue and orange circles represent the maximum 
and minimum envelopes in the ensemble prediction. The predictions are 
spread over a large range and though the prediction for the regular 
synoptic map has a large deviation, several of the standard deviation 
maps seem to have predicted the solar wind much closer. 
A possible reason for ``missing'' the two high speed streams in CR~2137 
could be the difficulty of modeling the complexity of the magnetic field 
configuration, during the late-ascending phase of the solar cycle, to which 
CR~2137 belong. It is noted that there were three ICMEs reported in the 
Richardonson/Cane catalog \citep{can03,ric10}; on May 26th (mean speed 
660 km/s), June 6th (mean speed 450 \kms) and June 8th (mean speed
430 km/s) during CR~2137 (2013). These are marked by the red vertical 
lines in the middle panel in Figure~\ref{fig:swspred}. The solar wind speeds  
before the start of ICMEs were about 600, 470 and 450~\kms, respectively. 
These values are hourly averages from the OMNI data of in situ solar wind
measurements but the results we presented here are the daily averages;
this, along with the fact that the mean speed of the ICMEs were comparable 
to the solar wind speed before and after the CME passage, implies that most 
of the effects of the CME must have been smoothed out.
Despite this, the model seem to have reproduced the solar wind 
modulations reasonably.

%
Figure~\ref{fig:swserr} shows the uncertainty estimates for the 
98~Monte-Carlo realizations for the three Carrington rotations 
selected for the present study. Here, the RMS errors between 
observed and predicted \sws$\;$using the CSSS  model for CR~2102 
is represented by the black dashed line, CR~2137 by the red 
dashed line and CR~2160 by the blue dotted line. The horizontal 
solid lines and the numbers within the brackets on the right 
hand top corner depict the mean RMS for the respective Carrington 
rotations indicated. It is noted that the errors, in general, 
tend to be larger during solar maximum as evident from the larger 
values seen for CR~2137. As clear from the figure, there is a 
large scatter for the RMS error for all the CRs studied and 
suggests that the errors in the arrival times of solar wind and 
other solar disturbances (e.~g. CMEs) are largely influenced by 
the errors (standard deviation) in the synoptic maps. 

%
%
\section{CONCLUDING REMARKS}
         \label{sec:discussion} 

The standard deviation maps (Figure~\ref{fig:hmi_maps}) represent 
the uncertainties in the preparation of the photospheric magnetic 
flux density synoptic maps from the magnetograms and do not represent 
the errors in the measurements of the magnetic field. This uncertainty 
is in addition to all other contributions such as instrument noise and
magnetic flux imbalance. Our aim in this paper is not to provide a method 
or a model with better predictive capability but to present a confidence level 
or an uncertainty estimate in the predicted \sws$\;$based on the uncertainties 
in the synoptic map used as boundary data to the model that predicted the 
\sws. This information (the uncertainty estimate), along with their origin 
(or source) is critical for devising methods to improve the prediction accuracies.
Given that such information (uncertainty estimates) is not usually provided 
when the synoptic maps are produced by various observatories (ground-based 
and spacecraft measurements) or when solar wind prediction is carried out, 
the relevance of the uncertainty estimate becomes all the more significant.
Our aim, in this paper, is to convey this point by demonstrating how the errors
(or uncertainties) in the boundary data (the synoptic maps) used in the models 
for solar prediction propagates and influence the accuracy of these predictions,
and to emphasize how important it is to incorporate this information (the
uncertainties) into future efforts to improve the prediction accuracy.

The photospheric magnetic flux density synoptic maps have been 
produced and used for decades for interpreting the various 
solar phenomena and observations. It is well established that 
the solar wind prediction is highly sensitive to the quality of 
these synoptic maps \citep[e.g.][]{arg00,arg10,ril14} that are
used as the inner boundary conditions of coronal models based on 
which the solar wind prediction have been made. Our use of 
magnetic vector field measurements gives us genuine observations 
of the radial flux distribution, which is not the case with the 
longitudinal field measurements usually employed in global 
coronal/heliospheric modeling. However, an estimate of
uncertainties in the construction of these maps have never 
been provided until \citet{ber14} produced the spatial standard 
deviation maps (\S~\ref{sec:method}). Similarly, a comprehensive 
estimate of uncertainties in the predictions of \sws$\;$\& HMF, 
locations of CH \& NL and other solar wind properties are 
also unavailable. Lack of such estimates results in poor 
understanding of the causes and incorrect identification of the 
sources of the discrepancies between predictions and observations, 
and thereby, inadequate mitigation of these factors.  

In this paper, we obtained systematic and reliable estimates of 
uncertainties of the coronal and solar wind properties predicted 
using the HMI photospheric flux density synoptic maps and the 
corresponding spatial standard deviation synoptic maps \citep{ber14}. 
For this, we computed the locations of the CHs (photospheric 
footpoints of open field regions) and NLs, and the FTEs and 
\sws$\;$using the Current Sheet Source Surface (CSSS) model 
\citep{zha95a,pod14,pod16} of the corona which takes the synoptic 
map as the inner boundary condition. We carried out the analysis
for three Carrington rotations CRs~2102 (3~--~30 October, 2010), 
2137 (14~May --~11 June, 2013) and 2160 (1~--~28 February, 2015),
representing the different phases of the solar cycle.
We compared the locations of CHs 
and NLs with the corresponding locations in the EUV synoptic maps 
for the same periods and those computed by the well--established 
PFSS model. The models and observations exhibit a close match, in 
general, as seen in Figures~\ref{fig:chnl1} -- \ref{fig:chnl3}.
A quantitative comparison of the \sws$\;$predicted by the CSSS 
model for these Carrington rotations with the correspsonding in 
situ observations of solar wind taken from the OMNI database was 
made by obtaining the RMS error as shown in Figure~\ref{fig:swserr}. 
We noted that there is considerable spread in the predicted \sws$\;$as
reflected in the RMS errors over the different standard deviation maps 
(MCRs) for a given Carrington rotation. Moreover, the RMS errors
are larger during CR~2137, a period during the solar maximum 
phase -- this is mainly due to the difficulty in modeling the complex 
magnetic field configuration during solar maximum as expected.  
The uncertainty in the solar wind prediction, on average, based on 
Figure~\ref{fig:swserr}, varied between 88 and 110~km/s, 
indicating a significant ambiguity between the slow and fast 
winds which has serious implications in space weather forecast. 

While the uncertainties originating due to the lack of far side 
information of the Sun, limitations of the model used  
calibration errors in the synoptic map construction, and the uncertainty 
in the propagation and arrival time of CMEs, a few to mention, are still 
significant factors in determining the prediction accuracy, the present 
analysis points out that the spread in the results, as shown here, is 
due to the spread in the boundary data values in the Monte Carlo 
simulations (especially at the poles).

The coronal models, the empirical relationship for solar wind 
prediction and the kinematic approach for the forward propagation 
of solar wind we employed here are simple and computationally 
inexpensive\footnote{
For the present work, the solar wind predictions 
using the 98~MCRs for each of the three Carrington rotations costed 
us about 100~days of CPU time (\S~\ref{sec:data}).}
but they form the 
basis of the current solar wind prediction schemes such as  the operational
WSA model. Therefore, though our model represent a static 
corona and does not handle transients (as already known to the 
scientific community), our results can be directly compared with 
those of the state-of-the-art model and other sophisticated space 
weather forecast models such as Enlil. Further, the present work 
indicates that (Figure~\ref{fig:swspred}) with appropriate corrections to 
the photospheric synoptic maps, the accuracy of solar wind prediction 
can be improved significantly. Moreover, the results presented here 
indicate the importance of ``ensemble forecast'' in improving the 
space weather forecast, as the results indicate that a suitable 
combination of the top performing models (lower values of RMSEs) 
can provide better and more accurate solar wind prediction. 

In order to obtain a better statistics and generalize our findings, 
we intent to expand the current study over longer periods of time. 
Since the near--Sun observations of Parker Solar Probe have 
already been released to the public, our near--Sun predictions can 
be better validated for optimizing the model performance, 
and, thereby, improve our near--Earth predictions as well.

%
%
\begin{acks} 
   Bala Poduval wishes to acknowledge Dr.~X.P.~Zhao for providing 
   her with the CSSS model and the many discussions that were 
   helpful in this work. \\
   This work used the Extreme Science and Engineering 
   Discovery Environment (XSEDE) \textit{Comet} 
   (\url{https://dl.acm.org/citation.cfm?id=2616540}) 
   at the the San Diego Supercomputer Center (SDSC) through a start 
   up allocation.
\end{acks}

%
%
\vspace{-20pt}

\bibliographystyle{spr-mp-sola}
\bibliography{ref_solar.bib}

\end{article}  
\end{document}